\documentclass{aa}
\usepackage{graphics}
\usepackage{natbib}
\bibpunct{(}{)}{;}{a}{}{,}

\DeclareRobustCommand{\ion}[2]{%
\relax\ifmmode
\ifx\testbx\f@series
{\mathbf{#1\,\mathsc{#2}}}\else
{\mathrm{#1\,\mathsc{#2}}}\fi
\else\textup{#1\,{\mdseries\textsc{#2}}}%
\fi}

\def\apjl{ApJ}%
\def\apjs{ApJS}%
\def\ao{Appl.~Opt.}%
\def\aap{A\&A}%
\title{A multi--scale study of infrared  and  radio emission from Scd galaxy M33}
\author{F. S. Tabatabaei\inst{1}, R. Beck\inst{1}, M. Krause\inst{1}, E. M. Berkhuijsen\inst{1}, R. Gehrz\inst{2}, K. D. Gordon\inst{3}, J. L. Hinz\inst{3}, R. Humphreys\inst{2}, K. McQuinn\inst{2}, E. Polomski\inst{2}, G. H. Rieke\inst{3}, C. E. Woodward\inst{2}}

\institute{Max-Planck Institut f\"ur Radioastronomie, Auf dem H\"ugel 69, 53121 Bonn, Germany   
\and School of Physics and Astronomy, University 
of
Minnesota, Minneapolis, MN 55455  
\and Steward Observatory, University of Arizona, 933 North Cherry Avenue, Tucson, AZ 85721
}

\begin{document}

\titlerunning{A multi--scale study of IR  and  radio emission from M33}
\authorrunning{Tabatabaei et al.}

\abstract
{}
{ We investigate the energy sources of the infrared\,(IR) emission and their relation to the radio continuum emission at various spatial scales within the Scd galaxy M33. }
{ We use the data at the $ Spitzer$ wavelengths of 24, 70, and 160\,$\mu$m, as well as  recent radio continuum maps at 3.6\,cm and 20\,cm observed with the 100--m Effelsberg telescope and VLA, respectively. We use the wavelet transform of these maps to a) separate the diffuse emission components from compact sources, b) compare the emission at different wavelengths, and c) study the radio--IR correlation  at various spatial scales. An H$\alpha$ map serves as a tracer of the star forming regions and as an indicator of the thermal radio  emission. }
{The bright HII regions affect the wavelet spectra causing dominant small scales or decreasing trends towards the larger scales.  The dominant scale of the 70\,$\mu$m emission is larger than that of the 24\,$\mu$m emission, while the 160\,$\mu$m emission shows a smooth wavelet spectrum. 
The radio and H$\alpha$ maps are well correlated with all 3 MIPS maps, although their correlations with the 160\,$\mu$m map are weaker. After subtracting the bright HII regions, the 24 and 70\,$\mu$m maps show weaker correlations with the 20\,cm map than with the 3.6\,cm map at most scales. We also find a strong correlation between the 3.6\,cm and H$\alpha$ emission at all scales. }
{Comparing the results with and without the bright HII regions, we conclude that the IR emission is influenced by young, massive stars increasingly with decreasing wavelength from 160 to 24\,$\mu$m. The radio--IR correlations indicate that the warm dust--thermal radio correlation is stronger than the cold dust--nonthermal radio correlation at scales smaller than 4\,kpc. A perfect 3.6\,cm--H$\alpha$ correlation implies that extinction has no significant effect on H$\alpha$ emitting structures. 
}
\maketitle

%

\section{Introduction}

One of the most important discoveries of the IRAS mission was the correlation between the IR and radio continuum luminosities for a sample of galaxies   \citep[e.g.,][]{Helou_etal_85,deJong_etal_85,Gavazzi_etal_86}. \cite{Beck_88} showed that this correlation also holds within several spiral galaxies including M31 and M33. The radio--IR correlation was explained by \cite{Helou_etal_85} and \cite{deJong_etal_85} as a direct and linear relationship between star formation and IR emission. On the other hand, there was concern that the cold dust emission, as a component of the far--infrared (FIR) emission, might not be directly linked to the young stellar population.  For the nearby galaxy M31, \cite{Hoernes_etal_98}, using IRAS data, found a good correlation between the emission of thermal--radio/warm dust and nonthermal--radio/cold dust with slightly different slopes. They explained the latter correlation by a coupling of the magnetic field to the gas mixed with the cold dust. Such a coupling was also considered by \cite{Niklas_97} as the origin of the global radio--FIR correlation. \cite{Hinz} found similar behavior in M33. 

The possibility that the relationship between the radio and IR emission might vary within galaxies \citep[e.g.,][]{Gordon_04} due to the range of individual conditions makes it necessary to repeat these kinds of  studies with improved data in the most nearby galaxies. Furthermore, it is uncertain  what component of a galaxy provides the energy that is absorbed and re--radiated in the IR \citep{Kennicutt_98}. For example, from the close correspondence between hydrogen recombination line emission and IR morphologies,
\cite{Devereux_etal_94b,Devereux_etal_96,Devereux_etal_97} and \cite{Jones_etal_02} argued that the FIR is powered predominantly by young O/B stars. \cite{Deul_89}, \cite{Walterbos_96}, and \cite{Hirashita_etal_03} argued that about half of the IR emission or more is due to dust heated by a diffuse interstellar radiation field that is not dominated by any particular type of star or star cluster. \cite{Sauvage_etal_92}  suggested that the relative role of young stars compared with the diffuse interstellar radiation field increases with later galaxy type.  
\begin{table}
\caption{Images of M33 used in this study }
\begin{tabular}{ l l l} 
\hline

Wavelength & Resolution & Telescope \\

\hline
20\,cm       &  $51\arcsec$ & VLA $^{1}$\\
20\,cm       &  $540\arcsec$ & Effelsberg $^{2}$\\
3.6\,cm      &  $84\arcsec$ & Effelsberg$^{1}$\\
160\,$\mu$m  &  $40\arcsec$ & Spitzer$^{3}$\\
70\,$\mu$m  &  $18\arcsec$ & Spitzer$^{3}$\\
24\,$\mu$m  &  $6\arcsec$ & Spitzer$^{3}$\\
6570\AA{}\,(H$\alpha$)   &  $2\arcsec$ (pixel size)  & KPNO$^{4}$ \\
\hline
\noalign {\medskip}
\multicolumn{3}{l}{$^{1}$ Tabatabaei et al. (in prep.)}\\
\multicolumn{3}{l}{$^{2}$ \cite{Fletcher}}\\
\multicolumn{3}{l}{$^{3}$ \cite{Hinz} and this paper }\\
\multicolumn{3}{l}{$^{4}$ \cite{Hoopes_et_al_97H}}\\
\end{tabular}
\end{table}

Another problem that hinders a better understanding of the radio--IR correlation is the separation of thermal and nonthermal components of the radio continuum emission using a constant nonthermal spectral index (the standard method). Although this assumption leads to a proper estimation for global studies, it cannot produce a realistic image of the nonthermal distribution in highly resolved and local studies within a galaxy, because it is unlikely that the nonthermal spectral index remains constant across a galaxy \citep{Fletcher_etal_04}.

\begin{figure*}
\begin{center}
\resizebox{15cm}{!}{\includegraphics*{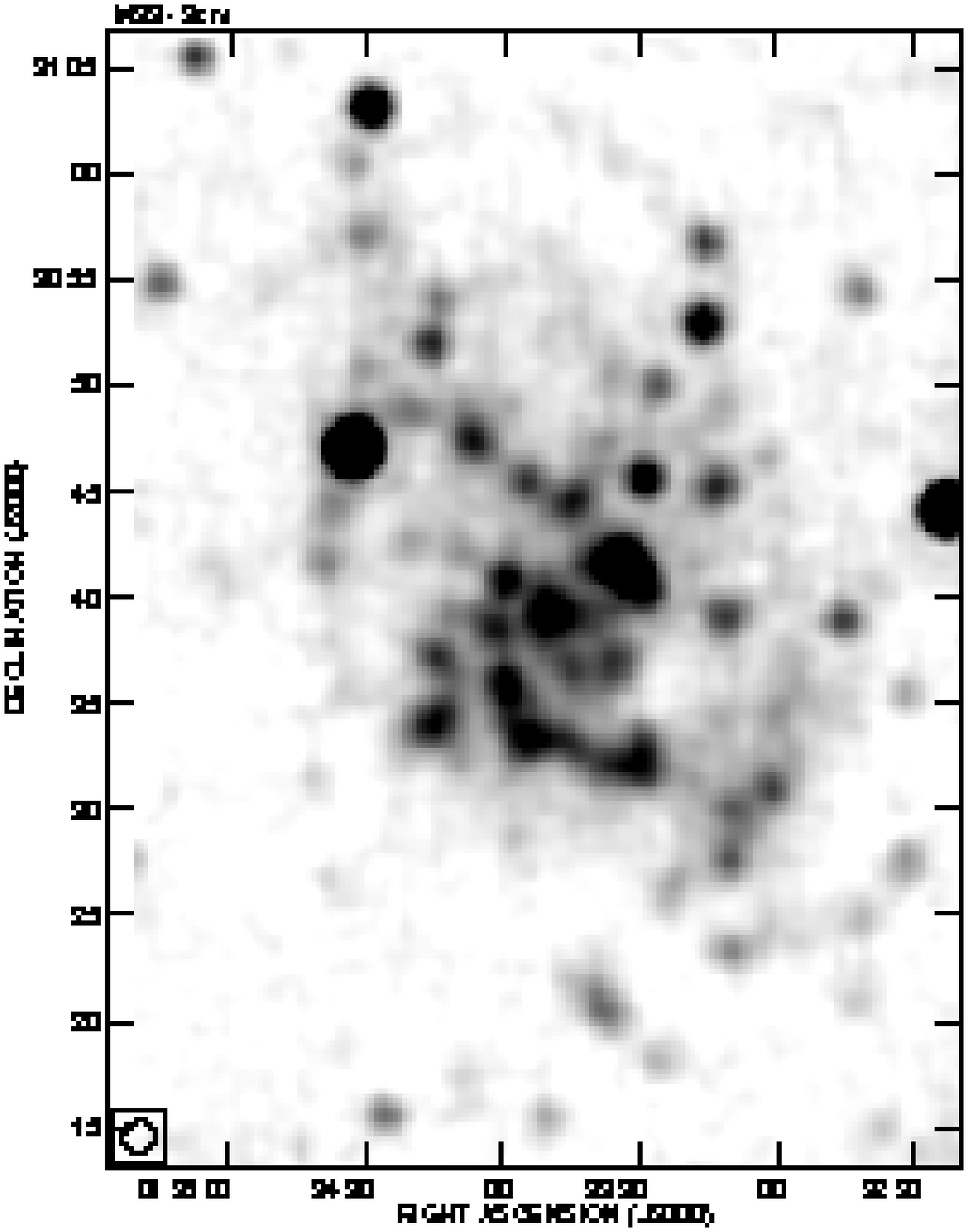}
\includegraphics*{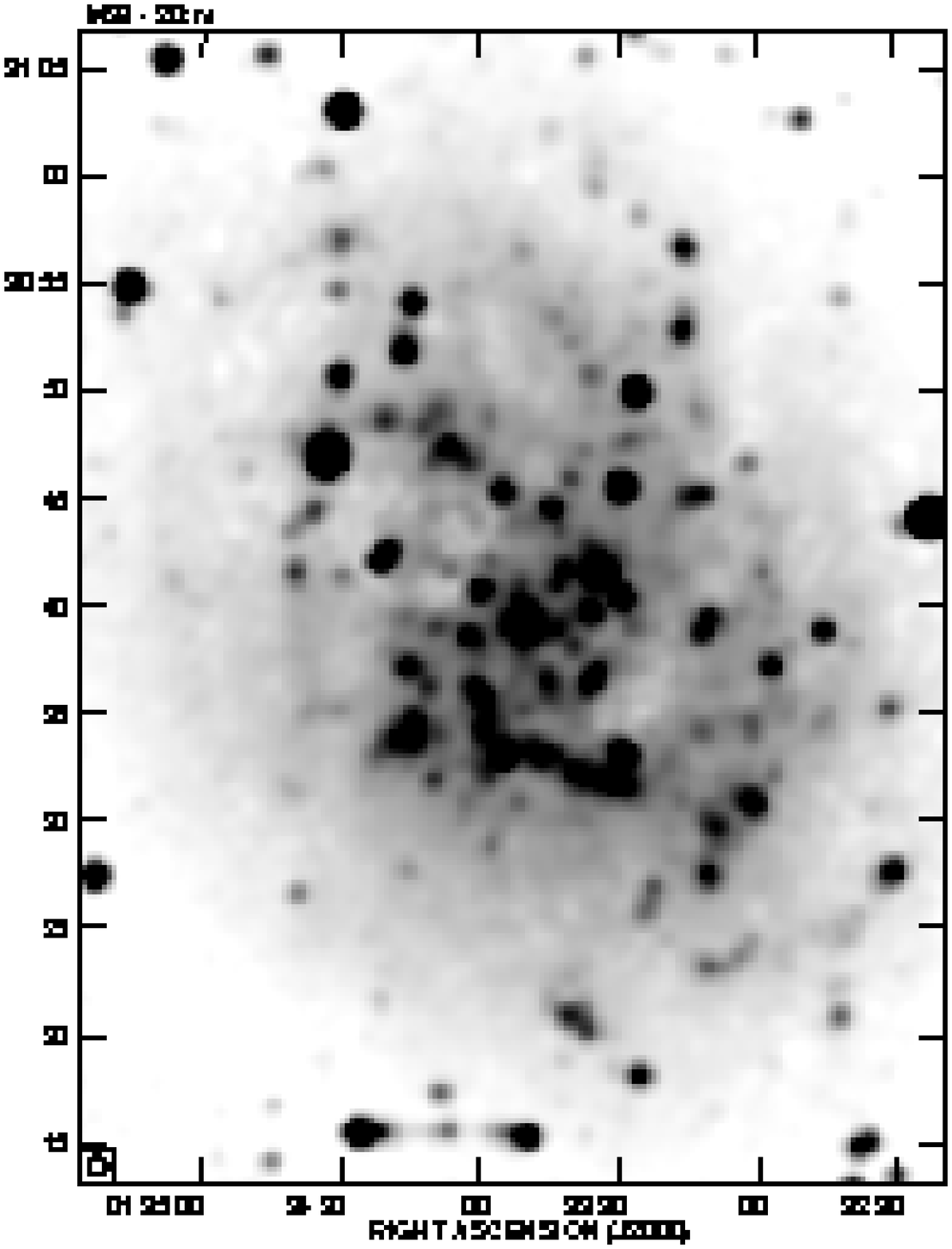}}
\caption[]{ The 3.6\,cm radio map of M33 (left panel) observed with the 100--m radio telescope in Effelsberg, and 
the combined 20\,cm radio map (right panel) from the VLA and Effelsberg observations. The half power beam widths (HPBWs) of $84\arcsec$ and $51\arcsec$, respectively, are shown in the lower left corners.  }
\end{center}
\end{figure*}

\begin{figure*}
\begin{center}
\resizebox{13.5cm}{!}{\includegraphics*{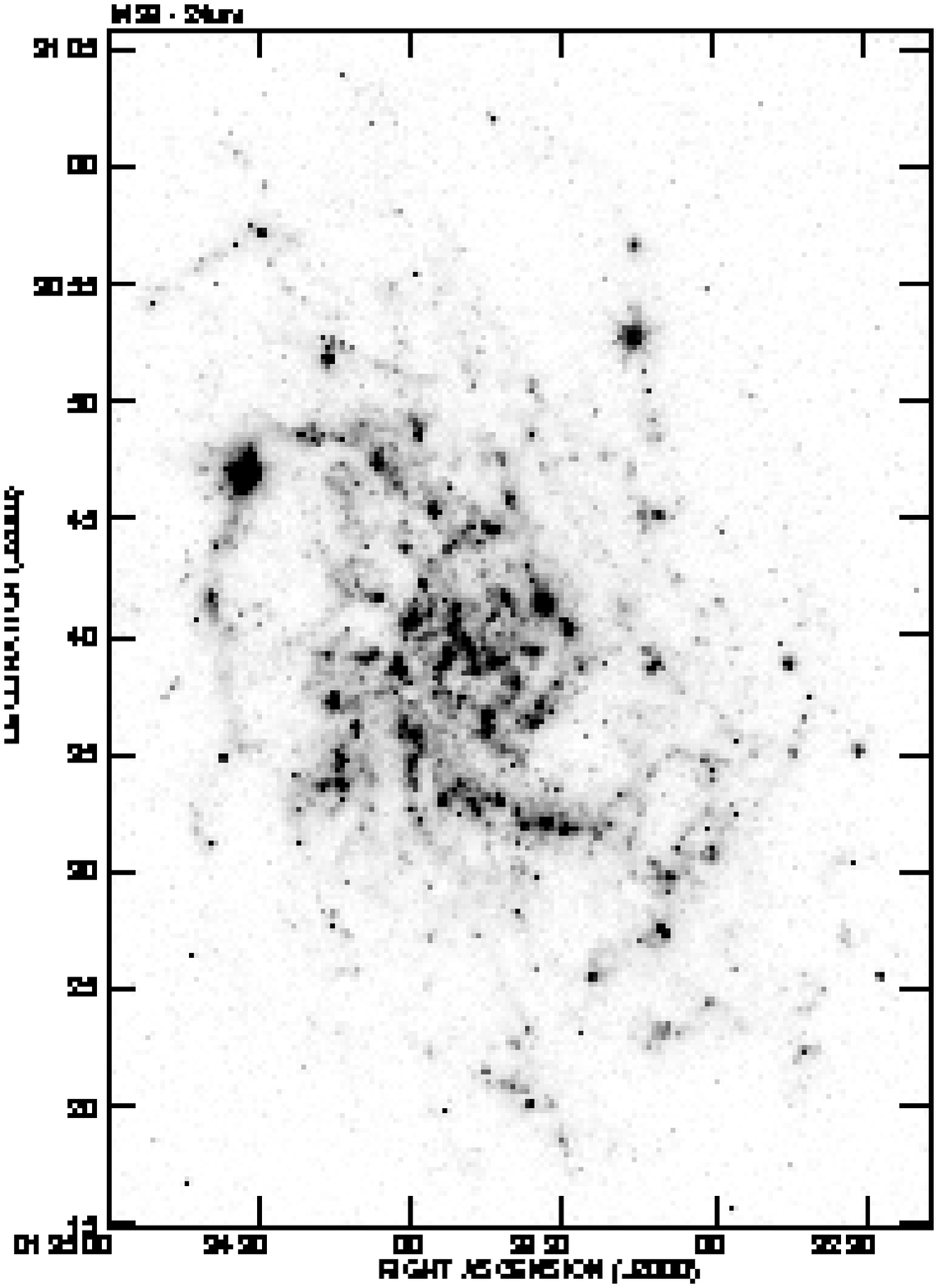}
\includegraphics*{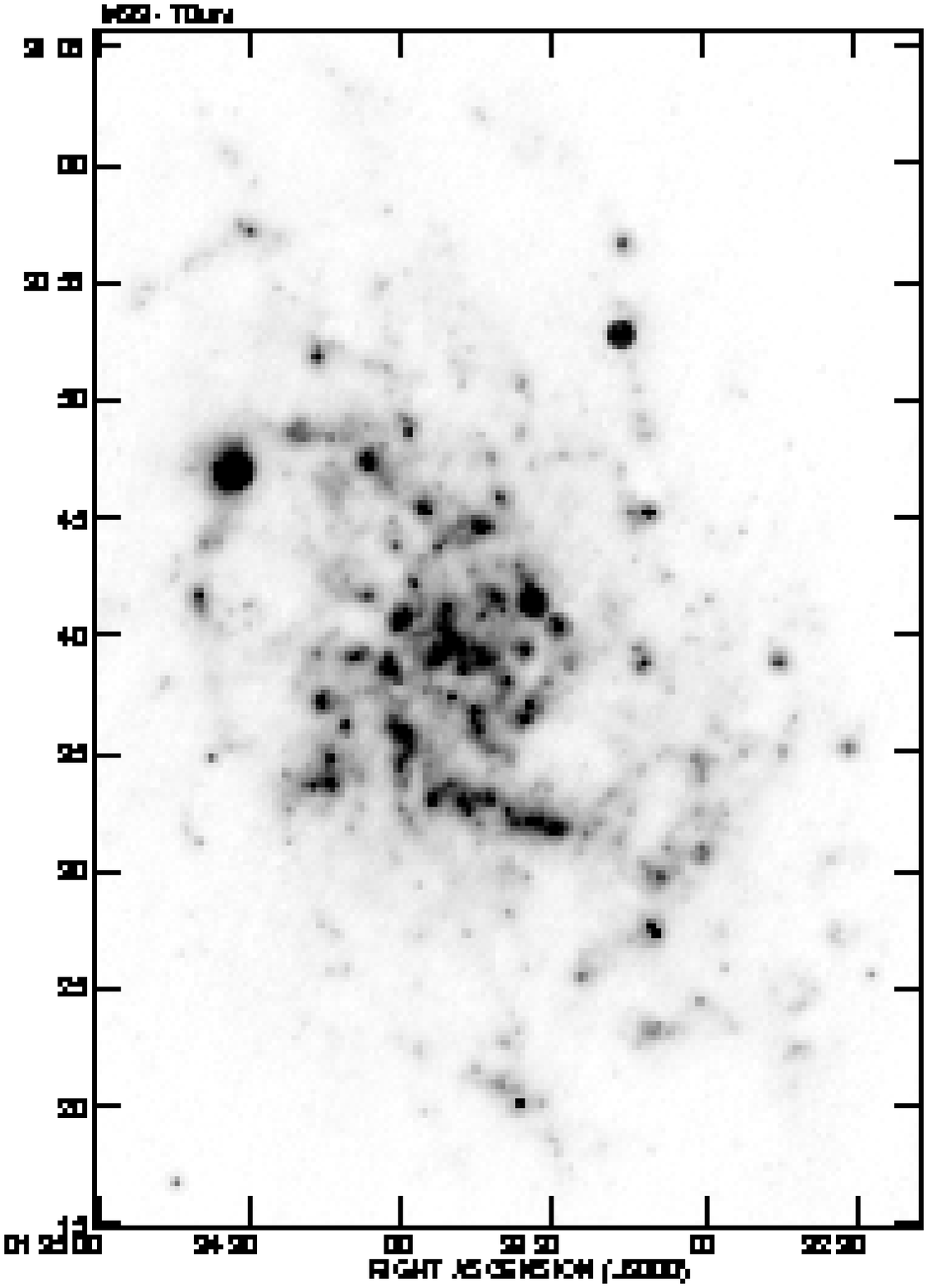}}
\resizebox{13.5cm}{!}{\includegraphics*{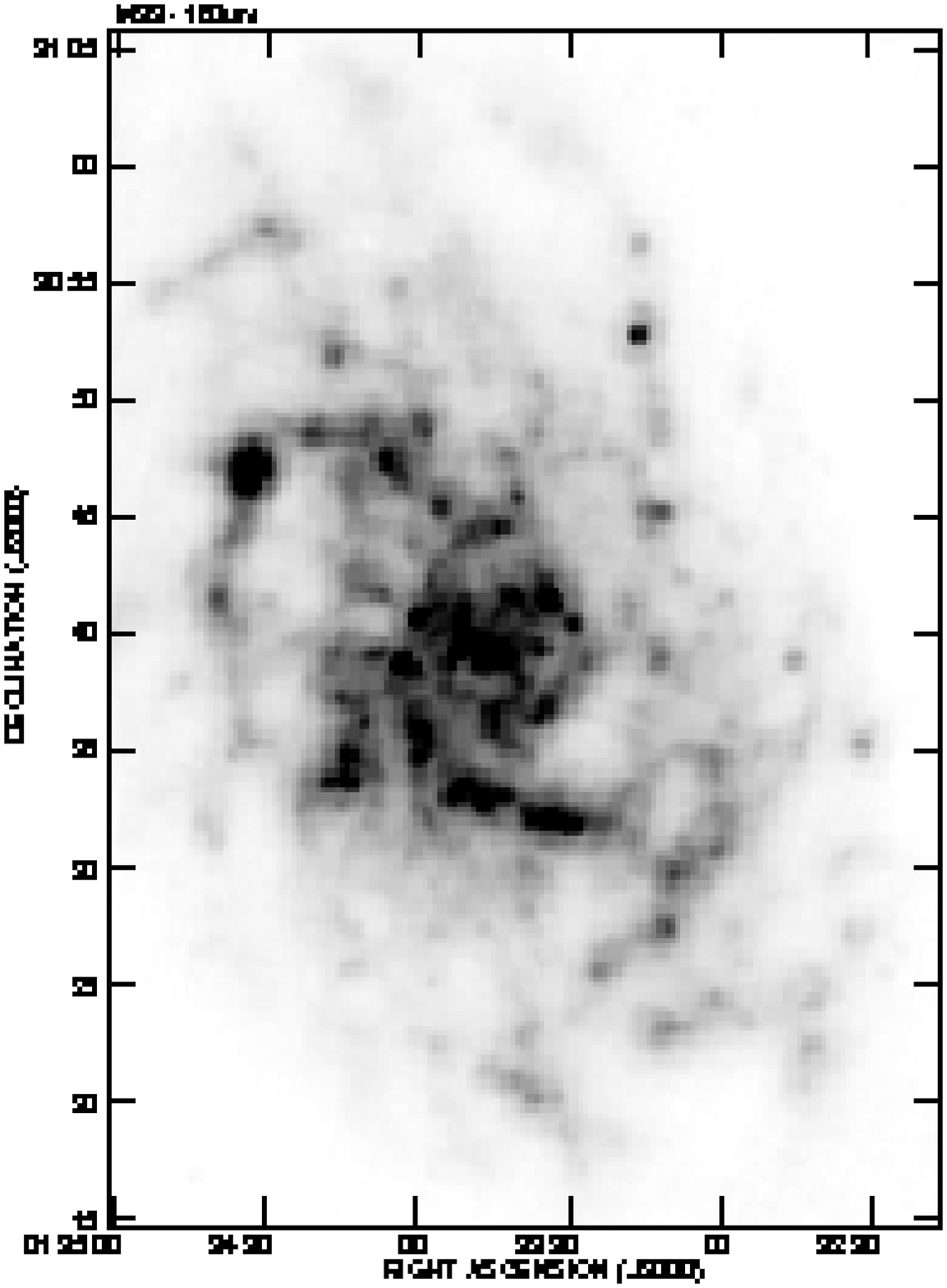}
\includegraphics*{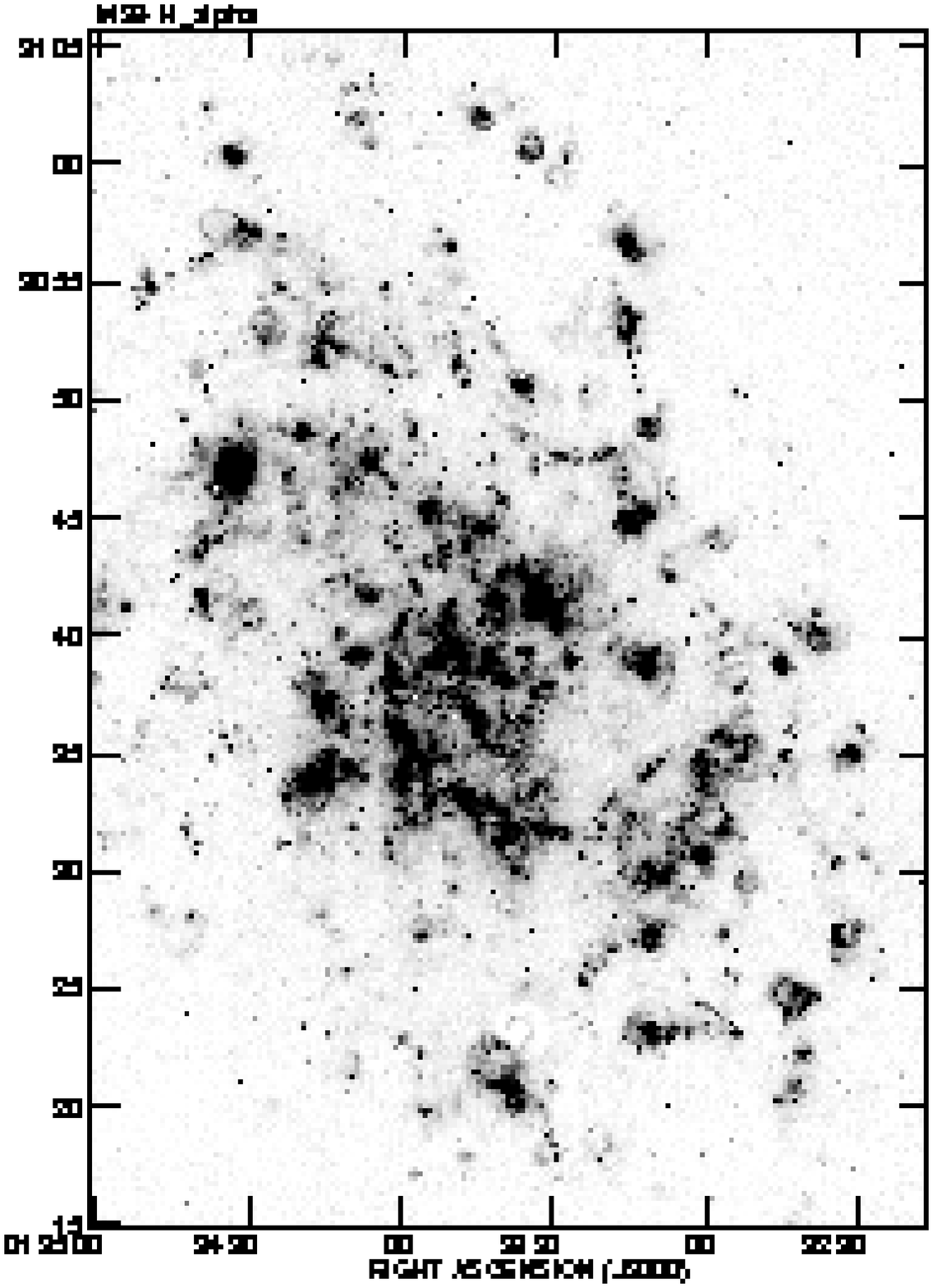}}
\caption[]{{\emph Top:} the Spitzer MIPS images of M33 at 24\,$\mu$m (left panel) and 70\,$\mu$m (right panel).
{\emph Bottom:} the Spitzer MIPS image at 160\,$\mu$m (left panel) and the H$\alpha$ map (right panel) from KPNO~\cite{Hoopes_et_al_97H}. The resolutions are given in Table 1.}
\end{center}
\end{figure*}

M33 (NGC\,598) is the nearest late--type spiral
galaxy, at a distance of 840\,kpc \citep{Freedman_etal_91} and is ideal to compare the morphologies of different components of such a galaxy. With an inclination of $i$\,=\,56$^{\circ}$ \citep{Regan_etal_94}, it is seen nearly face on. The central position of M33 given by \cite{devaucouleurs_81} is RA(2000)\,=\,$1^{h}33^{m}51.0^{s}$ and  DEC(2000)\,=\,$30^{\circ}39\arcmin37.0\arcsec$. The position angle of the major axis is  PA\,$\sim 23^{\circ}$ \citep{Deul}.

With the Multiband Imaging Photometer Spitzer \citep[MIPS,][]{Rieke} data at 24, 70, and 160\,$\mu$m, it is possible to interpret the morphology of M33.  \cite{Hinz} compared the first epoch data of MIPS with H$\alpha$ images and radio continuum data at 6\,cm, using a Fourier filtering technique to distinguish different emission components. We now have multiple epochs of MIPS data that provide a significantly higher--quality image of M33, for example in the suppressing of stripes along the direction of scan caused by slow response from the far infrared arrays. In addition, we use a wavelet analysis technique that \cite{Frick_etal_01} have shown to be more robust against noise than Fourier filtering. Furthermore, the wavelet technique provides more precise and easier analysis of the scale distribution of emission energy, especially at smaller scales of the emitting structures. We apply a 2D--wavelet transformation to separate the diffuse emission components from compact sources in MIPS mid-- and far--infrared (hereafter IR), radio (at 3.6 and 20\,cm), and H$\alpha$ images (Table 1).

The classical correlation between IR and radio emission has been at the whole galaxy level. However, such a correlation can be misleading when a bright, extended central region or an extended disk exists in the galactic image. This technique gives little information in the case of an anticorrelation on a specific scale \citep{Frick_etal_01}.  The `wavelet cross--correlation' introduced by \cite{Frick_etal_01} calculates the correlation coefficient as a function of the scale of the emitting regions.  \cite{Hughes_etal_06} applied this method to study the radio--FIR correlation in the Large Magellanic Cloud (LMC).

\cite{Hippelein} found evidence for a local radio--FIR correlation in the star forming regions of M33.
In this paper,  we investigate this correlation not only for the star forming regions but also for other structures within M33 at spatial scales between 0.4 and 4 kpc using the wavelet cross--correlation. Instead of the standard method to separate the thermal and nonthermal components of the radio continuum emission, we use the radio continuum data both at high and low frequency (3.6 and 20\,cm, respectively). We compare  our radio images to the H$\alpha$ image (which is taken as a tracer of the thermal free--free emission) to distinguish the components of the radio continuum emission.

We study the structural characteristics of the MIPS IR images using the 2D-wavelet transformation in Sect. 3. The distribution of the emission energy at both IR and radio regimes and the dominant spatial scale at each wavelength are discussed in Sect. 4. We show how  the different MIPS images are correlated at different scales and how the radio--IR correlation varies with components of the galaxy (e.g. gas clouds, spiral arms, extended central region and extended diffuse emission) in Sect. 5.  In Sec. 6, we use the H$\alpha$ map to probe the energy sources of IR and radio emission. Results and discussion are presented in Sect. 7. An overview of the preliminary results was presented in \cite{Tabatabaei_1}.

\section{Observations and data reduction}

Table 1 summarizes the data used in this work. 
The 3.6\,cm radio observations were made with the 100-m Effelsberg telescope\footnote{The 100--m telescope at Effelsberg is operated by the Max-Planck-Institut f\"ur Radioastronomie (MPIfR) on behalf on the Max--Planck--Gesellscahft.} during several periods (18 nights) between August 2005 and March 2006. The reduction was done in the NOD2 data reduction system ~\citep{Haslam}.  The r.m.s. noise  after 
 combination \citep{Emerson_etal_88} of 36 coverages is $\sim$\,220 $\mu$Jy/beam. 
We also observed M33 at 20\,cm with the B--band VLA\footnote{The VLA is a facility of the National Radio Astronomy Observatory. The NRAO is operated by Associated Universities, Inc., under contract with the National Science Foundation.} D-array during 5 nights in November 2005 and January 2006 (from 06 to 08-11-05,  13-11-05, and 06-01-06). The reduction, calibration, and mozaicing of 12 fields were accomplished using the standard AIPS programs. The VLA interferometric data will miss much of the extended emission of the galaxy. For example, \cite{Viallefond} showed that the WSRT interferometric map at 1.4 GHz accounted for only about 16\% of the total emission. Therefore, we combined the VLA map with the new Effelsberg 20\,cm map \citep{Fletcher} to recover the emission from extended structures in M33. The r.m.s. noise of the final map is $\sim$\,180\,$\mu$Jy/beam. Detailed descriptions of the observations and data reduction of the radio data will be given in Tabatabaei et al. (in prep.). Both radio maps are shown in Fig. 1.

\begin{figure*}
\resizebox{\hsize}{!}{\includegraphics*{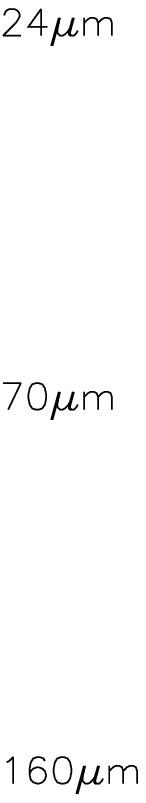}
\includegraphics*{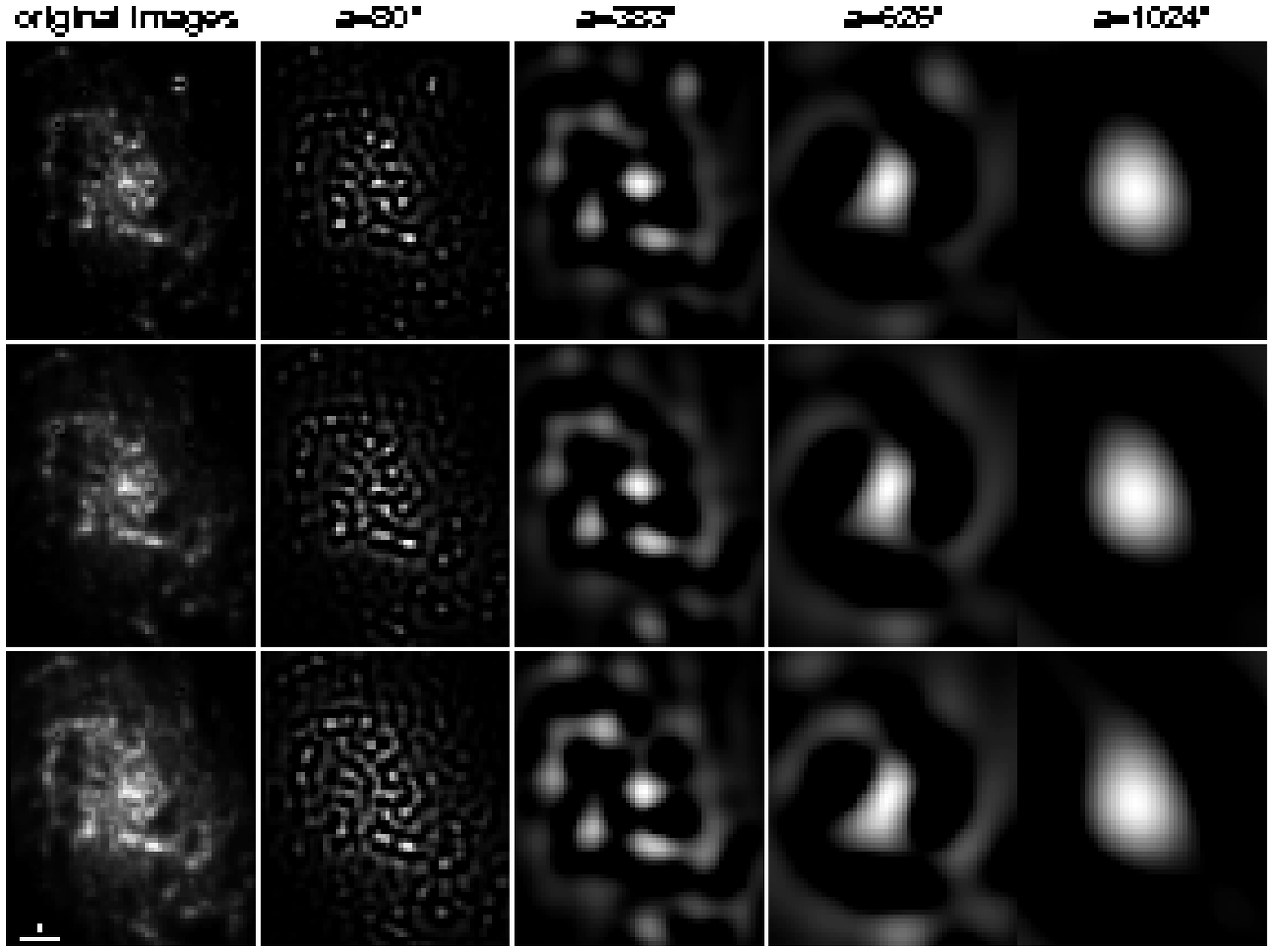}}
\caption[]{MIPS maps of M33 at $40\arcsec$ resolution (first column) and their wavelet decompositions for 4 different scales: $80\arcsec$ (second column), $383\arcsec$ (third column), $626\arcsec$ (forth column), and $1024\arcsec$ (fifth column). At the distance of 840 kpc, $1\arcsec$ is equivalent to 4\,pc. The maps at 24, 70, and 160\,$\mu$m are shown from top to bottom. Before the decomposition, huge HII regions like NGC\,604 were subtracted from the original images  (Sect. 4). Maps are shown in RA--DEC coordinate system and centered at the center of the galaxy. 
The field size is $34\arcmin$\,$\times$\,$41\arcmin$.}
\end{figure*}

M33 was mapped in the IR by MIPS \citep{Rieke} four times on 29/30
December 2003, 3 February 2005, 5 September 2005, and 9/10 January
2006.  Each observation consisted of medium-rate scan maps with 1/2
array cross-scan offsets and covering the full extent of M33.  The
basic data reduction was performed using the MIPS instrument team Data
Analysis Tool versions 3.02-3.04 \citep{Gordon_05}.  At 24\,$\mu$m extra steps were carried out to
improve the images including readout offset correction, array averaged
background subtraction (using a low order polynomial fit to each leg,
with the region including M33 excluded from this fit), and exclusion
of the first five images in each scan leg due to boost frame transients.
At 70 and 160\,$\mu$m, the extra processing step was a pixel dependent
background subtraction for each map (using a low order polynomial fit,
with the region including M33 excluded from this fit).  The background
subtraction should not have removed real M33 emission as the scan legs
are nearly parallel to the minor axis resulting in the background
regions being far above and below M33.  

The 24\,$\mu$m image used consisted of just the 9/10 January 2006
observations as the depth reached in a single mapping was sufficient
for this work.  The 70\,$\mu$m image used was a combination of the
last three observations as the 29/30 December 2003 70\,$\mu$m map
suffered from significant instrumental residuals.  These instrumental
residuals were much reduced when the operating parameters of the
70\,$\mu$m array were changed after the first M33 map was made and
before the second.  The 160\,$\mu$m image used consisted of a
combination of all four maps.  The combination of multiple maps of M33
taken with different scan mirror angles results in a significant
suppression of residual instrumental signatures (seen mainly as low
level streaking along the scan mirror direction).  The images used in
this work have exposure times of $\sim$100, 120, and 36
seconds/pixel for 24, 70, and 160\,$\mu$m, respectively.

The H$\alpha$ observations by \cite{Hoopes_et_al_97H} were carried out on the 0.6 meter Burrell--Schmidt telescope at the Kitt Peak National Observatory, providing a $68\arcmin$\,$\times$\,$68\arcmin$ field of view. The MIPS and H$\alpha$ images are shown in Fig. 2.

To obtain a proper comparison with the 3.6\,cm radio  map, all images of M33 were convolved to the angular resolution of $84\arcsec$. For higher angular resolution studies, maps at  $18\arcsec$, $40\arcsec$, and $51\arcsec$ were also made. The PSF~(point spread function) of the MIPS data is not Gaussian, in contrast to that of the radio data. Thus, convolutions of the MIPS images were made using custom kernels created using Fast
Fourier Transforms (FFTs) to account for the detailed structure of the
MIPS PSFs. 
Details of the kernel creation can be found in Gordon et al. (2007, in
prep.). After convolution, the maps were normalized in grid size, rotation and reference coordinates. Then, they were cut to a common field of view ($34\arcmin$\,$\times$\,$41\arcmin$ in RA and DEC, respectively).

In the following sections, we discuss the results with and without bright compact sources at each resolution to investigate how these sources affect
the energy distribution at each wavelength and
the correlations between wavelengths.
After convolving to each resolution, the bright sources, common to all images, were subtracted   using  Gaussian fits including baselevels (`Ozmapax' program in the NOD2 data reduction system). For instance, the giant HII complexes NGC604 and NGC595 were subtracted at all resolutions. We also subtracted strong steep--spectrum background radio sources from the 20\,cm image. Detailed descriptions of the source subtraction at each resolution are given in Sect. 4.   

\begin{table*}
\caption{Source subtraction thresholds S($\lambda$) and number of subtracted sources at different spatial resolutions. 
 }
\begin{tabular}{c c c c c c c} 
\hline

Resolution  & Subtracted sources & S(24\,$\mu$m) & S(70\,$\mu$m) & S(160\,$\mu$m) & S(20\,cm) & S(3.6\,cm) \\
arcsec & \# & $\mu$Jy/arcsec$^2$ & $\mu$Jy/arcsec$^2$ & $\mu$Jy/arcsec$^2$& $\mu$Jy/beam & $\mu$Jy/beam \\

\hline
18 & 40 & 460 & 2650 & -- & -- & --\\
51 & 15 & 115 & 1570 & 2865 & 7190 & --\\
84 & 11 & 85 & 900 &  2340& 8260& 5780\\
\hline
\end{tabular}
\end{table*}

\section{Wavelet analysis of IR emission}

Wavelet analysis is based on a spatial--scale decomposition using the convolution of the data with a family of self--similar basic functions that depend on the scale and location of the structure. Like the Fourier
transformation, the wavelet transformation includes oscillatory functions; however, in the latter case these functions rapidly decay towards infinity. As a result, \cite{Frick_etal_01} show that the wavelet method is more resistant to noise and the smoother spectra allow better determination of the true frequency structure. The cross--correlation of wavelet spectra lets us compare the structures of different images systematically as a function of scale.

The wavelet coefficients of a 2D continuous wavelet transform are given by:

\begin{equation}
W(a,{\bf x})=\frac{1}{a^{\kappa}} \int_{-\infty}^{+\infty}  f(\bf x')\psi^{\ast}(\frac{{\bf x'-x}}{{\it a}}){\it d}{\bf x},
\end{equation} 
\noindent
where $\psi({\bf x})$ is the analysing wavelet, ${\bf x} = (x,y)$, $f({\bf x})$ is a two--dimensional function (an image), and {\it a} and $\kappa$ are the scale and normalization parameters, respectively, (the $^{\ast}$ symbol denotes the complex conjugate).
The above transformation decomposes an image into `maps' of different scale. In each map, structures with the chosen scale are prominent as they have higher coefficients than those with smaller or larger scales. 
To obtain a good separation of scales and to find the scale of dominant structures in M33, we use the `Pet--Hat' wavelet that was introduced by \cite{Frick_etal_01} and applied there to NGC\,6946.
It is defined in Fourier space by the formula:

\begin{equation}
\psi(\bf k)=\left \{ \begin{array}{ll}
{cos^2(\frac{\pi}{2}log_{2}\frac{k}{2\pi})}  & \,\,\,\,\,\pi \le \bf k \le 4\pi \\ 
0 &   \pi > \bf k  \,\,\,\, \textrm{or} \,\,\,\,  \bf k > 4\pi , %
\end{array} \right.
\end{equation}
where ${\bf k}$ is the wavevector and $k=\vert \bf k \vert$.

\begin{figure}
\resizebox{\hsize}{!}{\includegraphics*{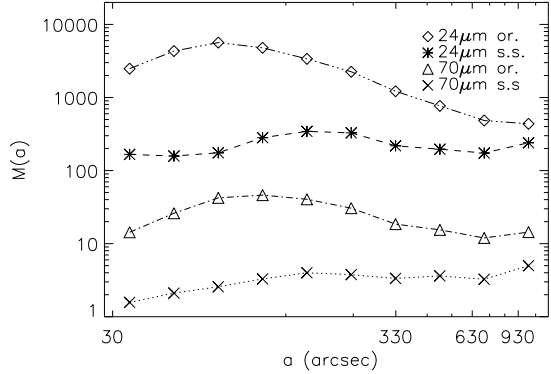}}
\caption[]{The wavelet spectra of the 24 and 70 $\mu$m images before (or.) and after (s.s.) subtraction of the same sources from the two images at  $18\arcsec$ resolution. The data points correspond to the scales 34, 50, 73, 107, 155, 226, 329, 479, 697, $1015\arcsec$. The spectra are shown in arbitrary units.}
\end{figure}

We decomposed the Spitzer images into 10 different scales {\it a}  to compare the morphologies at 24, 70, and 160\,$\mu$m  at each scale\footnote{To have both physically meaningful results and a sufficiently large number of independent points, the scale {\it a} varies between a minimum of about twice the resolution and a maximum of about half of the image size, $a < 1100\arcsec$,  for all images studied in this paper.}. Fig. 3 shows the original maps and the decomposed maps at 4 scales. The original 24 and 70\,$\mu$m maps were smoothed to the MIPS 160\,$\mu$m
PSF with FWHM~(full width half maximum)\,=\,$40\arcsec$ before decomposition.  The maps at scale {\it a}\,=\,$80\arcsec$ (0.3\,kpc) show the smallest detectable emitting structures. Most of the morphological differences among the MIPS images are found at this scale. At scale {\it a}\,=\,$383\arcsec$  (1.5\,kpc), the prominent structures are spiral arms and the center of the galaxy. The central extended region is more pronounced at scale {\it a}\,=\,$628\arcsec$ (2.5\,kpc). The emission emerges from a diffuse structure at scale {\it a}\,=\,$1024\arcsec$ (4\,kpc), and it is not possible to distinguish the arm--structure anymore. This structure can be identified as an underlying diffuse disk with a general radial decrease in intensity. The similarity of the 4\,kpc maps at different wavelengths indicates that the large scale diffuse emission has the same structure at different mid-- and far--infrared wavelengths.

At the smallest scale, the emission emerges from spot--like features aligned along filaments with the  width of the scale. At 24\,$\mu$m the spots, corresponding to HII regions, contain most of the energy at this scale (see Sect. 6). Diffuse filaments are more pronounced at 160\,$\mu$m . As shown in the next section, the fraction of the energy at this scale at 160\,$\mu$m is less than that at 24\,$\mu$m.    
 The situation in the 70\,$\mu$m map is in between the 24 and 160\,$\mu$m maps.
It seems that the star forming regions provide most of the energy of the 24 and 70\,$\mu$m emission, if the spots correspond to these regions, as is discussed in the following sections.

\begin{figure*}
\resizebox{\hsize}{!}{\includegraphics*{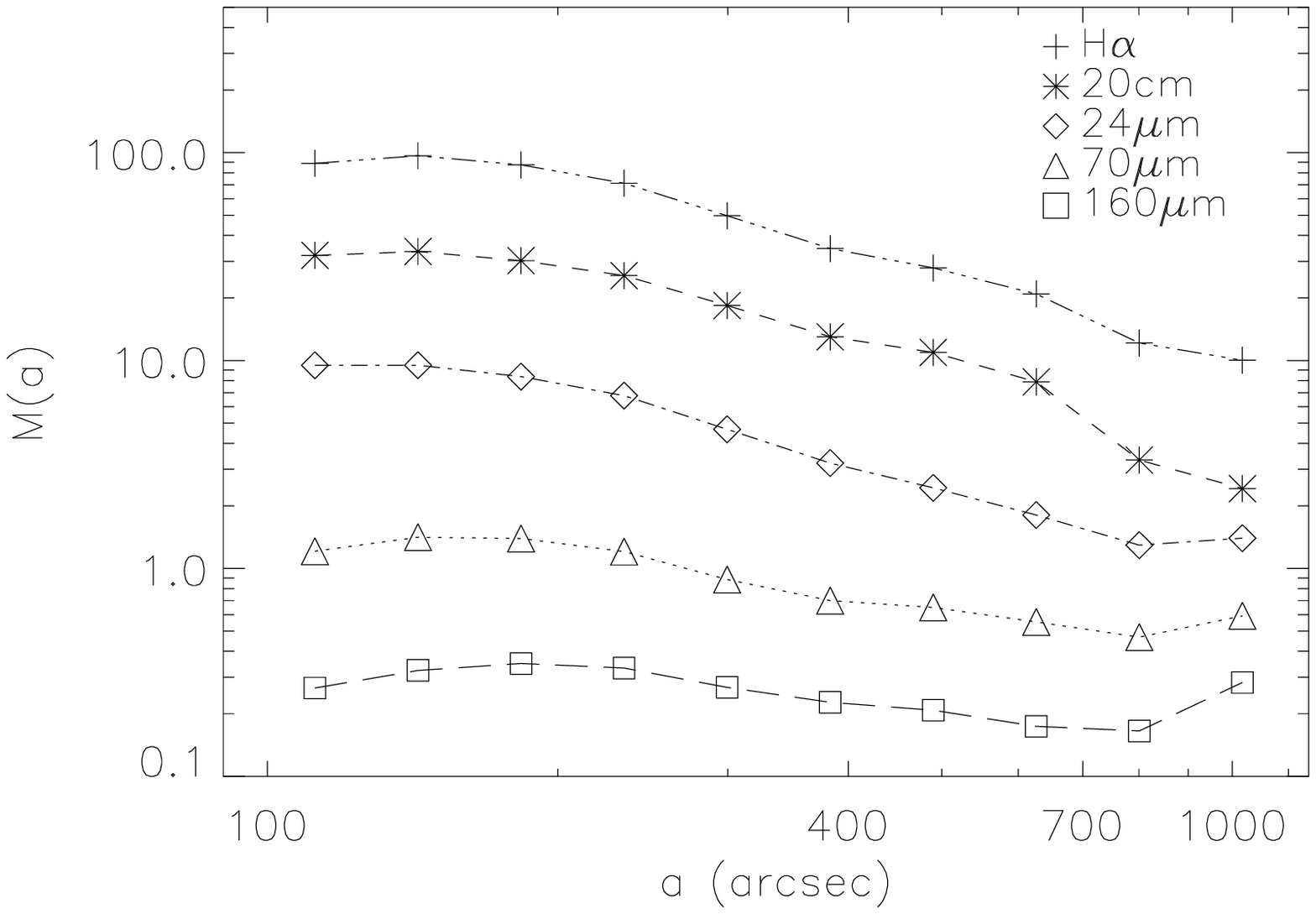}
\includegraphics*{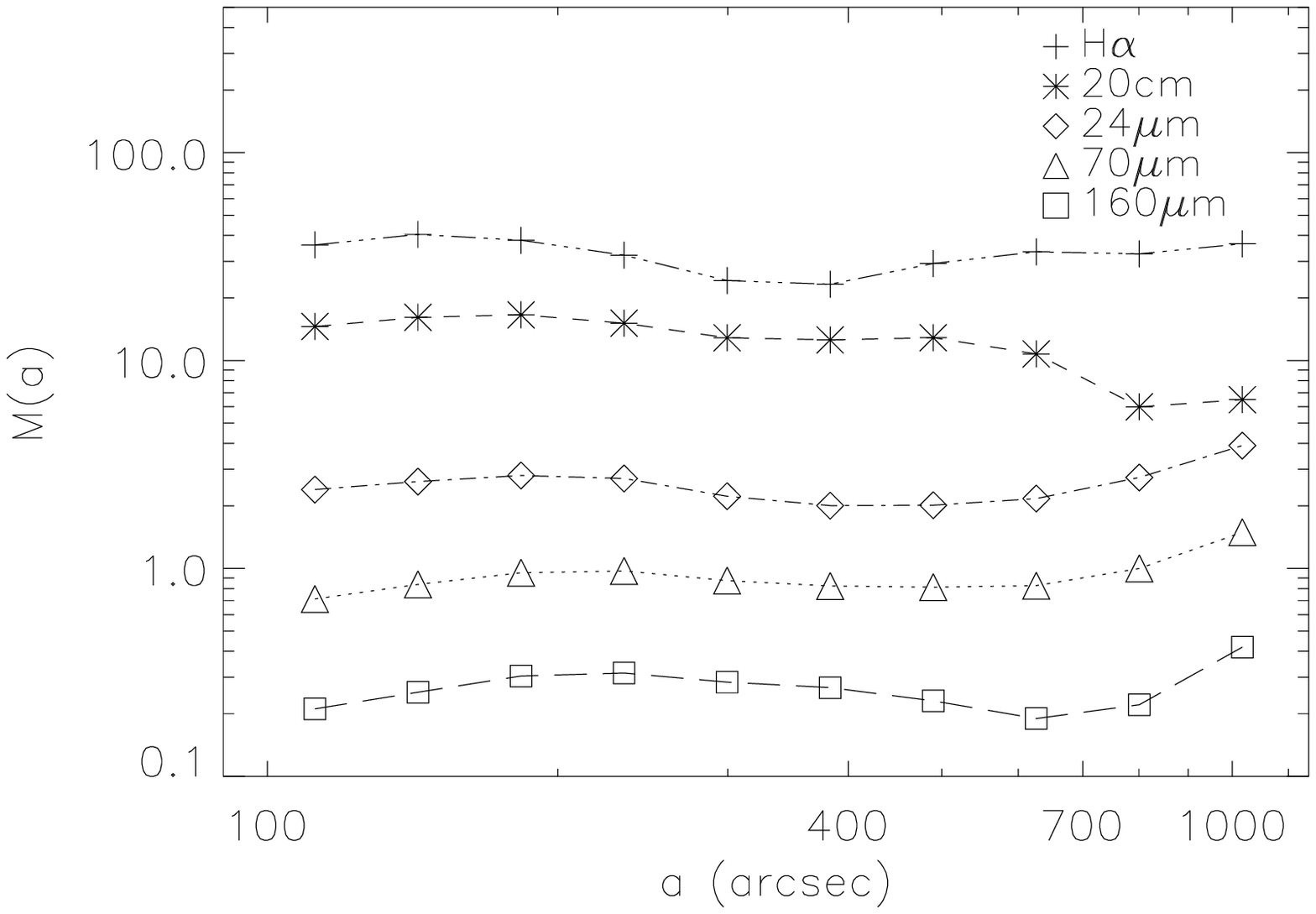}}
\caption[]{The wavelet spectra of the 20\,cm radio and IR images at $51\arcsec$ resolution before (left) and after (right) subtraction of the same sources. For comparison, the wavelet spectrum of the H$\alpha$ emission is also plotted. The data points correspond to the scales 112, 143, 183, 234, 299, 383, 490, 626, 800, $1024\arcsec$. The spectra are shown in arbitrary units. }
\end{figure*}

\begin{figure*}
\resizebox{\hsize}{!}{\includegraphics*{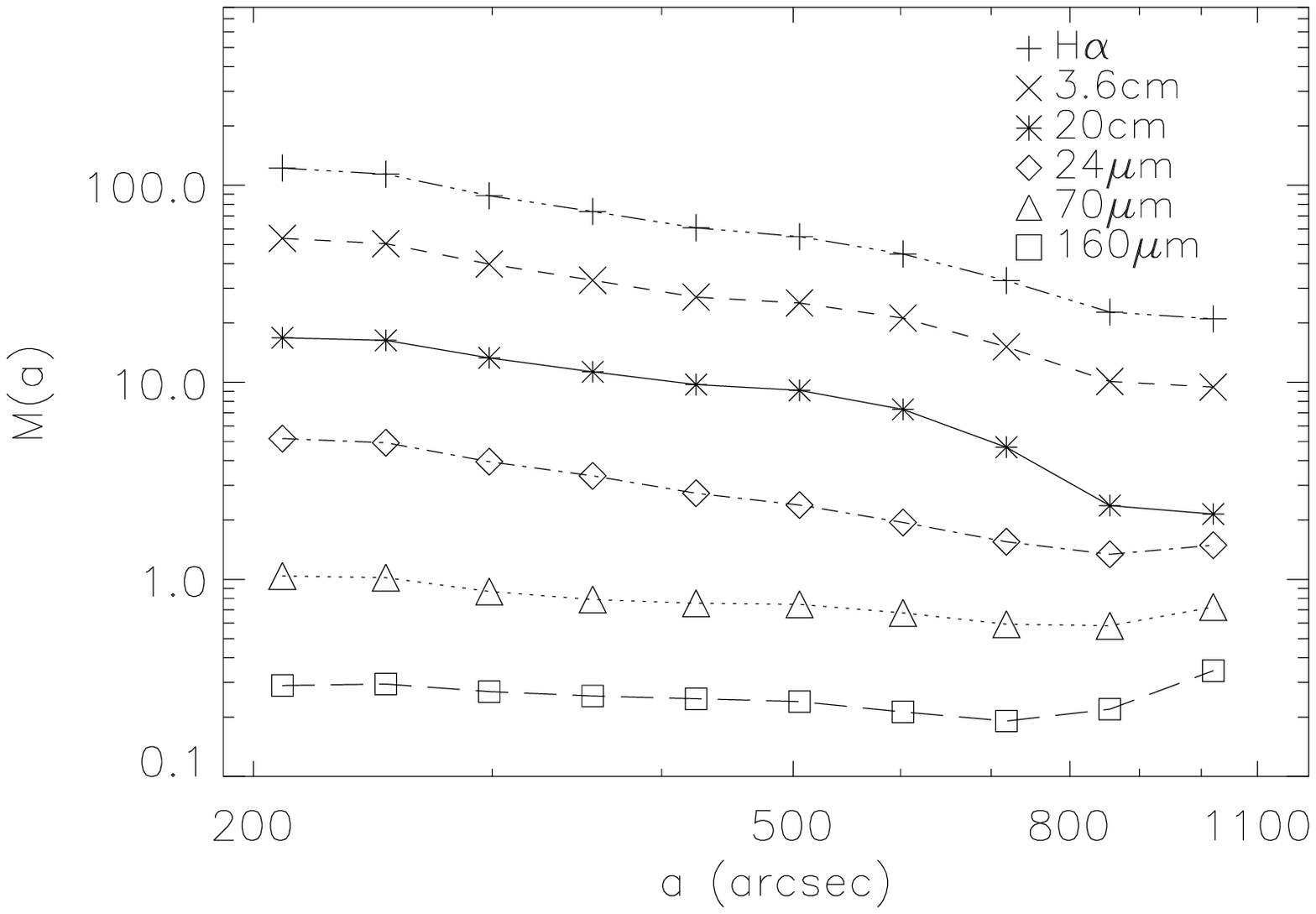}
\includegraphics*{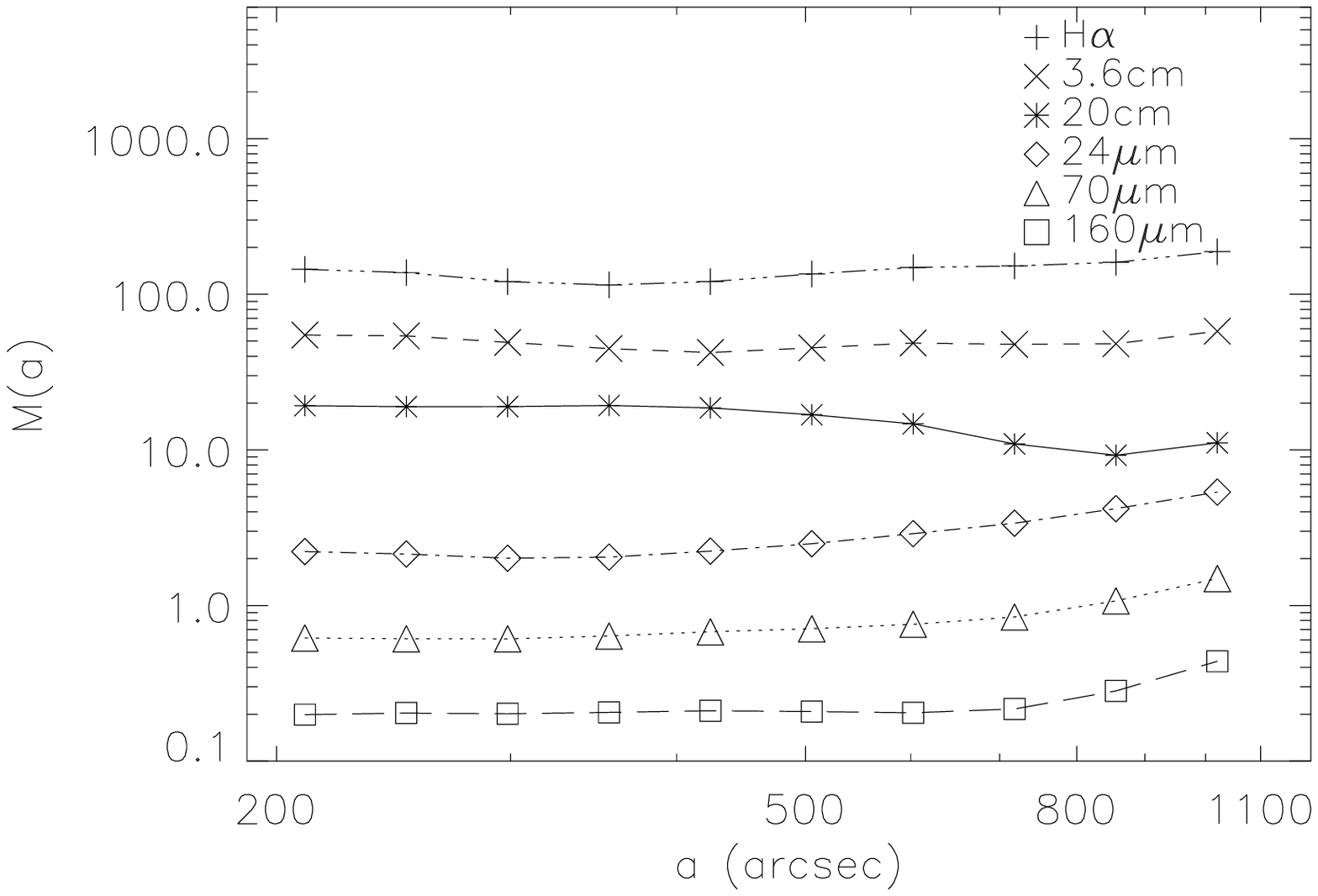}}
\caption[]{The wavelet spectra of the 3.6\,cm radio and IR images at $84\arcsec$  resolution before (left) and after (right) subtraction of the same sources. For comparison, the spectrum of the H$\alpha$ emission is also plotted. The data points correspond to the scales 210, 250, 298, 356, 424, 505, 602, 718, 856, $1020\arcsec$. The spectra are shown in arbitrary units.}
\end{figure*}

\section{Spectral characteristics of IR and radio maps}

In this section, we demonstrate how the wavelet spectra can be used to investigate the scaling properties of the emission. This spectrum is defined as the energy in the wavelet coefficients of scale {\it a} \citep{Frick_etal_01}\,: 

\begin{equation}
M(a) = \int_{-\infty}^{+\infty} \int_{-\infty}^{+\infty}  \vert W(a,{\bf x})\vert ^{2} d\bf x.
\end{equation}

After smoothing the 24\,$\mu$m map to the MIPS 70\,$\mu$m PSF with FWHM\,=\,$18\arcsec$, about 40 bright sources (mostly corresponding to HII regions) were subtracted from both maps (equivalent to removing sources with fluxes higher than the lower limits, S($\lambda$), given in Table 2.). Fig. 4 shows the wavelet spectra, M({\it a}), of the 24\,$\mu$m and 70\,$\mu$m maps before and after source subtraction. Clearly, the small scales  are the dominant scales before source subtraction. The scale at which the wavelet energy is maximum is $\sim 70\arcsec$ (280\,pc) at 24\,$\mu$m and $\sim$\,$110\arcsec$ (440\,pc) at 70\,$\mu$m. After source subtraction, the spectra become flatter. The larger size of the dominant scale at 70\,$\mu$m caused by these sources (by a factor of $\sim$\,1.6 at this resolution) indicates a slightly more extended distribution of dust grains emitting at 70\,$\mu$m than 24\,$\mu$m in the vicinity of star forming regions. Moreover, the ratio of the maximum to minimum energy in the original spectrum at 24\,$\mu$m is larger (by a factor of 3) than that at 70\,$\mu$m. This indicates that either star forming regions provide more energy for the 24\,$\mu$m emission, or the large--scale diffuse emission is stronger at 70\,$\mu$m than 24\,$\mu$m.

At the next angular resolution, $51\arcsec$, the HPBW~(half power beam width) of the 20\,cm radio map, 15 bright sources (HII regions) visible at all maps were subtracted from the 20\,cm map and the smoothed 24, 70, and 160\,$\mu$m IR maps. In addition to these sources, 9 background radio sources  plus the supernova remnants SN1, SN2, and SN3 \citep{viallefond_et_al_98} were subtracted from the 20\,cm radio map.  
The wavelet spectra of the MIPS and 20\,cm radio maps at this resolution are plotted in Fig. 5. The H$\alpha$ spectrum is also given for comparison. The 24 and 70\,$\mu$m maps show a smoothed version of their distributions in Fig. 4 (the linear smoothing factor is $\sim$\,3 between  Figs. 4 and 5). 
However, the effect of the sources can still be seen by comparing the left panel with the source subtracted spectra  in the right panel. 

The 160\,$\mu$m map is hardly influenced by the smoothing, as its original resolution is $40\arcsec$ (the smoothing factor is $\sim$\,1.3). The 160\,$\mu$m spectrum is more similar to the 70\,$\mu$m spectrum than to the other spectra. 
It seems that the compact sources have less effect on the energy distribution at 160\,$\mu$m, because the smallest scale is not the dominant scale. Hence, there is no important change in the spectrum after the source subtraction. The energy shows an increase at the scale of complexes of dust and gas clouds ($\sim$\,$250\arcsec$ or 1\,kpc), then a second increase in transition to the large--scale structures or  diffuse dust emission.

The spectrum of the 20\,cm  radio image is also  dominated by bright sources. There is a maximum at the scale  a\,=\,$140\arcsec$, then a decrease towards the larger scales with a slope of -0.9.  However,  a flat spectrum remains for the scales less than the size of the central extended region ($\sim$\,600$\arcsec$) after the source subtraction.

The spectra of all the maps at the resolution of the 3.6\,cm data (HPBW\,=\,$84\arcsec$)\footnote{Gaussian PSFs with FWHM\,=\,$51\arcsec$ and $84\arcsec$ are considered to convolve the IR maps to the angular resolutions of the 20 and 3.6\,cm radio maps, respectively.} are shown in Fig. 6 (left panel). The spectra after subtracting the 11 brightest HII regions are plotted in the right panel. Again, because of these sources, the small scales are dominant at 3.6\,cm. As shown in comparing the left and right panels, they also are seen in all the infrared bands, strongly at 24\,$\mu$m and much more weakly at 160\,$\mu$m. Here, the smoothing effect is seen in all 3 MIPS wavelet spectra. 

Before source subtraction, the 20 and 3.6\,cm spectra are similar to each other and to the H$\alpha$ spectrum up to the scale of the central extended region, $\sim$\,$600\arcsec$ (2.5\,kpc).  All three spectra become steeper between {\it a}\,$\sim$\,$600\arcsec$ and {\it a}\,$\sim$\,$850\arcsec$, which means that the wavelet energy from structures with scales of about half of the size of the galaxy ($\sim$\,900$\arcsec$ or 3.7\,kpc) is not as significant as that from smaller structures. It seems that a minimum in the wavelet spectra of the radio and H$\alpha$ emission, as was also shown in NGC\,6946 \citep[see Fig. 7 in][]{Frick_etal_01}, is a characteristic of this scale. However, while this decrease disappears in the 3.6\,cm and H$\alpha$ spectra after source subtraction, it remains in the 20\,cm spectrum. This implies that besides the bright HII regions there are sources of nonthermal emission within the spiral arms and central extended region which emitt significantly at 20\,cm.\\

To estimate how much of the wavelet energy $M(a)$ is provided by the subtracted sources, we consider the following definition:
\begin{equation}
\Delta (a) \equiv \frac{M_{or.}(a)-M_{s.s.}(a)}{M_{or.}(a)} ,
\end{equation}
where $M_{or.}(a)$ and $M_{s.s.}(a)$ represent the wavelet energy before and after source subtraction. 
Table 3 shows the fraction of energy produced by the 11 HII regions in IR and at 3.6\,cm for 3 different scales (at $84\arcsec$ resolution). The corresponding fractions at 20\,cm are not shown in this table, as some nonthermal radio sources were also subtracted at this wavelength. The emission at 24\,$\mu$m has the largest energy fraction $\Delta(a) $ at all scales. The smallest $\Delta(a) $ occurs at 160\,$\mu$m. This indicates that HII regions play a more important role in providing energy for dust emission at 24\,$\mu$m than at 160\,$\mu$m. 

\begin{table}
\begin{center}
\caption{Fractions of the wavelet energy provided by the 11 brightest HII regions at different scales (Eq. 4).}
\begin{tabular}{ l  l  l  l} 
\hline

$\lambda$ & $\Delta (210\arcsec)$ & $\Delta (505\arcsec)$ & $\Delta (1020\arcsec)$ \\

\hline
24\,$\mu$m  &  0.94 & 0.81 & 0.35\\
70\,$\mu$m  &  0.80 & 0.57 & 0.14\\
160\,$\mu$m  &  0.52 & 0.33 & 0.11\\
3.6\,cm  & 0.86  & 0.69 & 0.16\\
\hline
\end{tabular}
\end{center} 
\end{table}

We estimate  the uncertainties of the wavelet energies of each map by making a noise map with the distribution and amplitude of the real noise in the corresponding observed map. A linear combination of uniform and Gaussian distributions simulates the real noise in each observed image. The wavelet spectra of the derived noise maps were obtained using Eq. 3. The resulting wavelet energies of the noise maps are taken as  the uncertainties of the wavelet energies of the observed maps at different scales. The estimated values are at least three orders of magnitude less than the wavelet energies of the observed maps. Therefore, the results from our wavelet spectrum analysis are not substantially affected by the noise in the maps.

\begin{figure}
\resizebox{\hsize}{!}{\includegraphics*{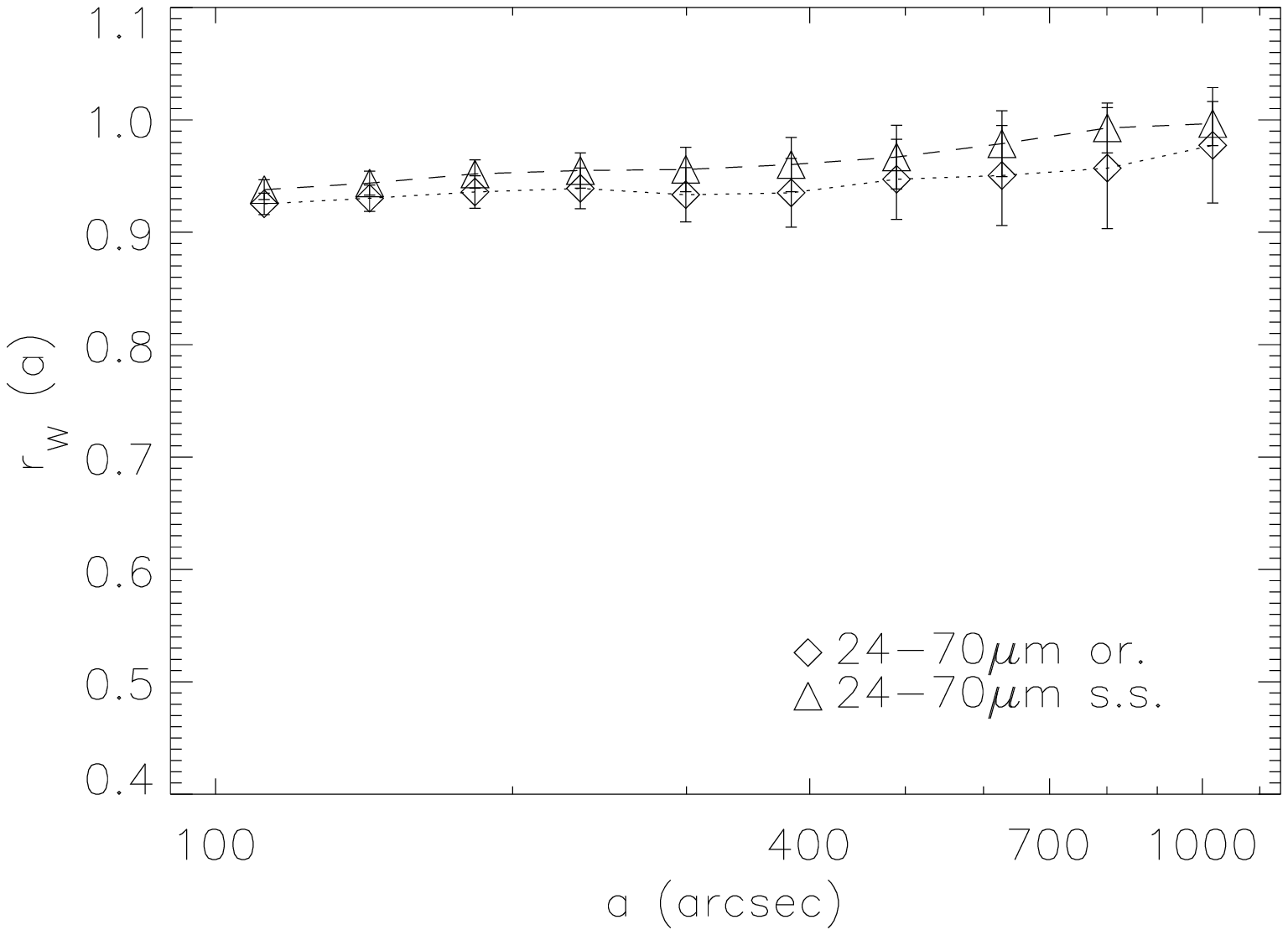}}
\resizebox{\hsize}{!}{\includegraphics*{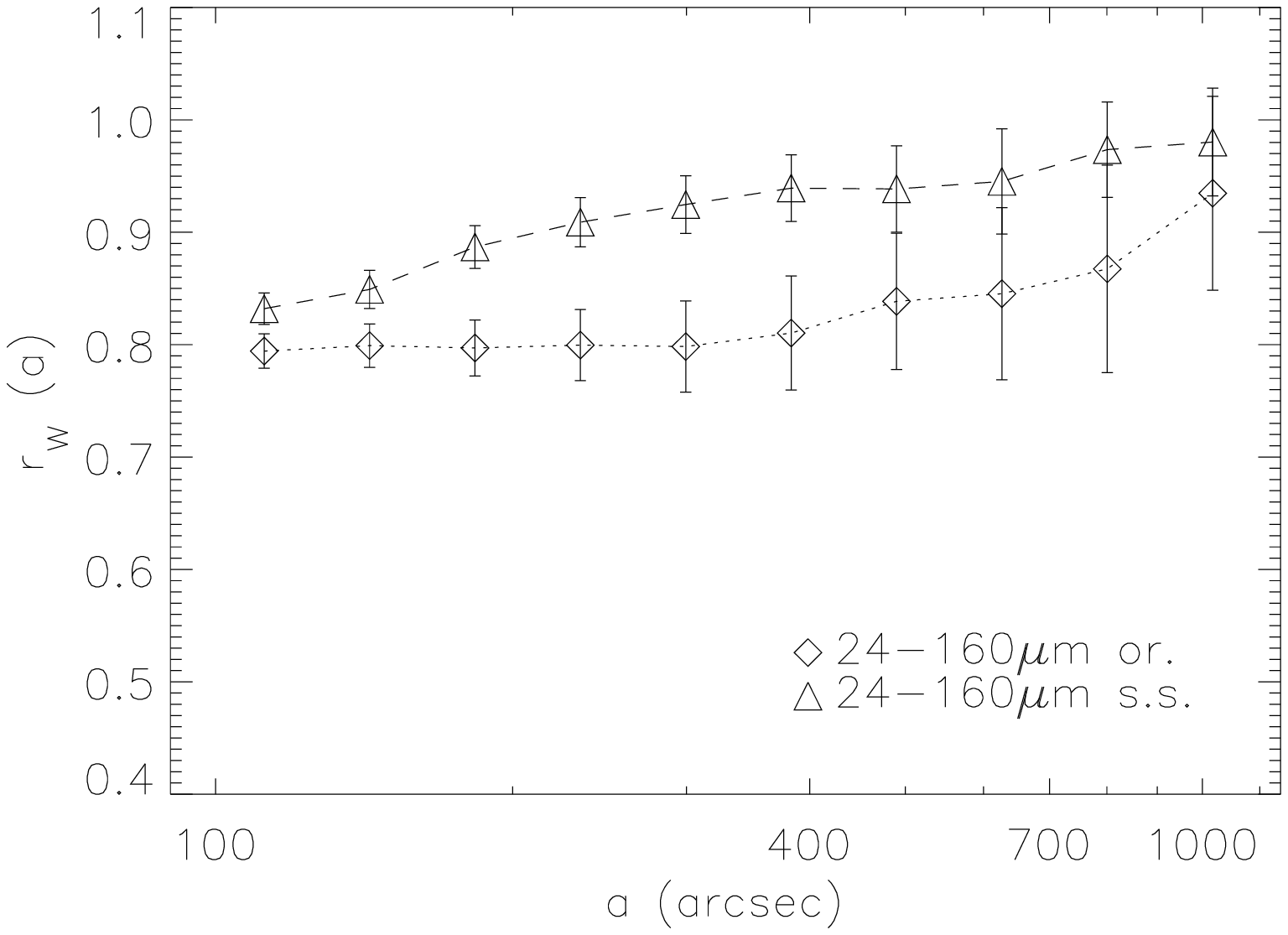}}
\resizebox{\hsize}{!}{\includegraphics*{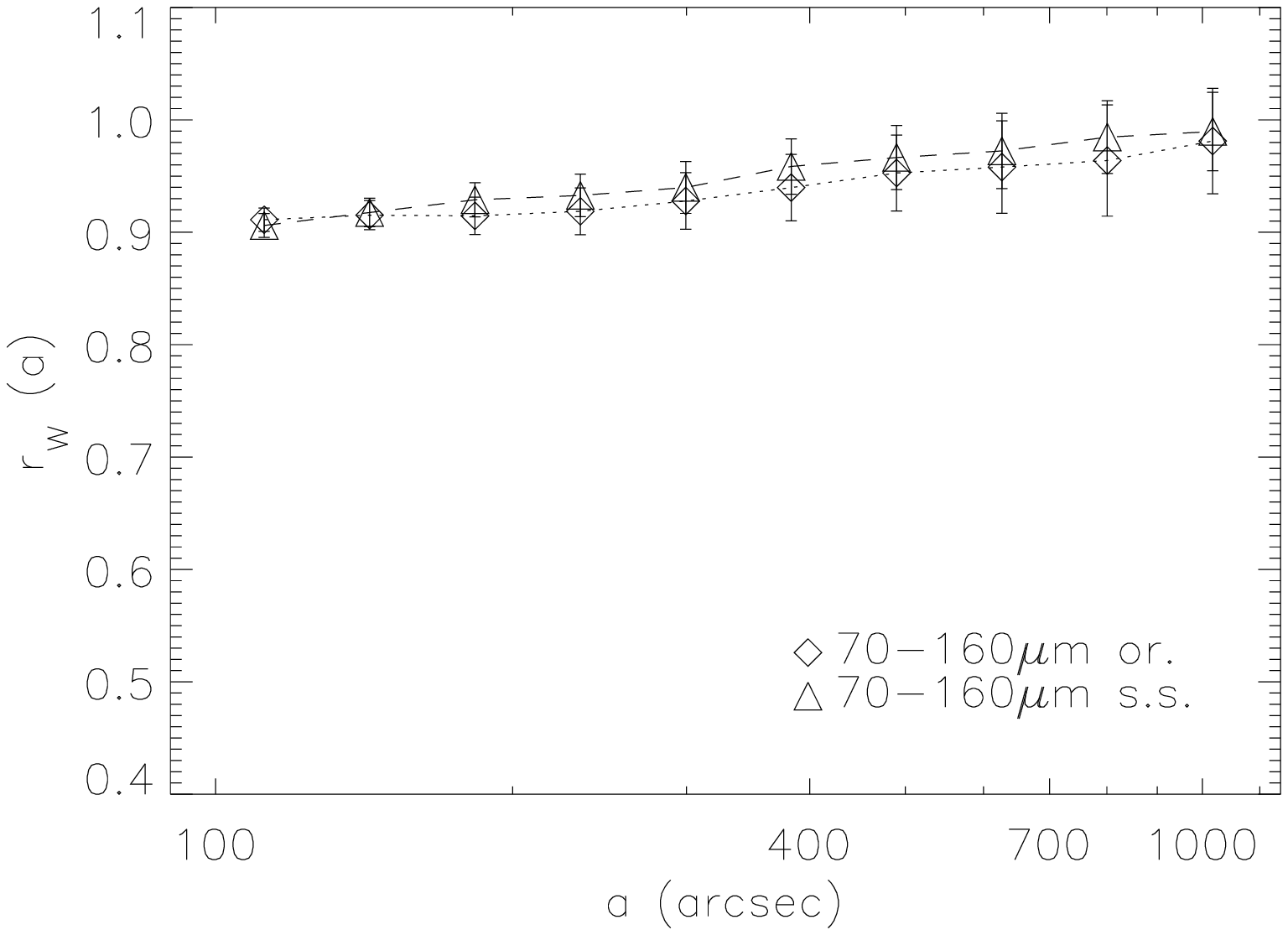}}
\caption[]{The cross--correlation between 24 and 70\,$\mu$m (top), 24 and 160\,$\mu$m (middle), and 70 and 160\,$\mu$m (bottom) images at $51\arcsec$ resolution before (or.) and after (s.s.) source subtraction.  }
\end{figure}

\section{Wavelet cross--correlations}

A useful method to compare images at different wavelengths is the wavelet cross-correlation. The wavelet cross-correlation coefficient at scale {\it a} is defined as:

\begin{equation}
r_{w}(a)=\frac{\int \int W_{1}(a,{\bf x})~W^{\ast}_{2}(a,{\bf x}) d{\bf x}}{[M_{1}(a)M_{2}(a)]^{1/2}},
\end{equation}
 where the subscripts refer to two images of the same size and linear resolution. 
The value of $r_{w}$ varies between -1 (total anticorrelation) and +1 (total correlation). Plotting $r_{w}$ against scale shows how well structures at different scales are correlated in intensity and location. 
The error is estimated by the degree of correlation and the number of independent points, n:

\begin{equation}
\Delta r_{w}(a)= \frac{\sqrt{1-r^{2}_{w}}}{\sqrt{n-2}},
\end{equation} 
where, $ n$\,=\,2.13\,$(\frac{L}{a})^2,  $ and $L$ is the size of the maps.

 First, we examine the cross--correlations between the 3 MIPS maps of M33 (Fig.7). 
There are significant correlations between each pair of the MIPS maps at different scales within 0.4\,$<$\,{\it a}\,$<$\,4\,kpc, as $r_{w} > 0.75$ \citep{Frick_etal_01}. These scales include gas clouds, spiral arms, and the central extended region. Comparing the plots, the 24--160\,$\mu$m correlation is weaker than the two other correlations, especially at scales smaller than the spiral arms ({\it a}\,$<$\,$400\arcsec$). This is due to significantly different contributions of HII regions in providing energy for dust emission at 24 and 160\,$\mu$m (see Table 3), as the 24--160\,$\mu$m correlation coefficients increase after subtracting these sources. This indicates that HII regions influence the correlations at scales larger than their sizes (the width of the huge HII region NGC604 is about $100\arcsec$ or 0.4\,kpc).

\begin{figure}
\resizebox{\hsize}{!}{\includegraphics*{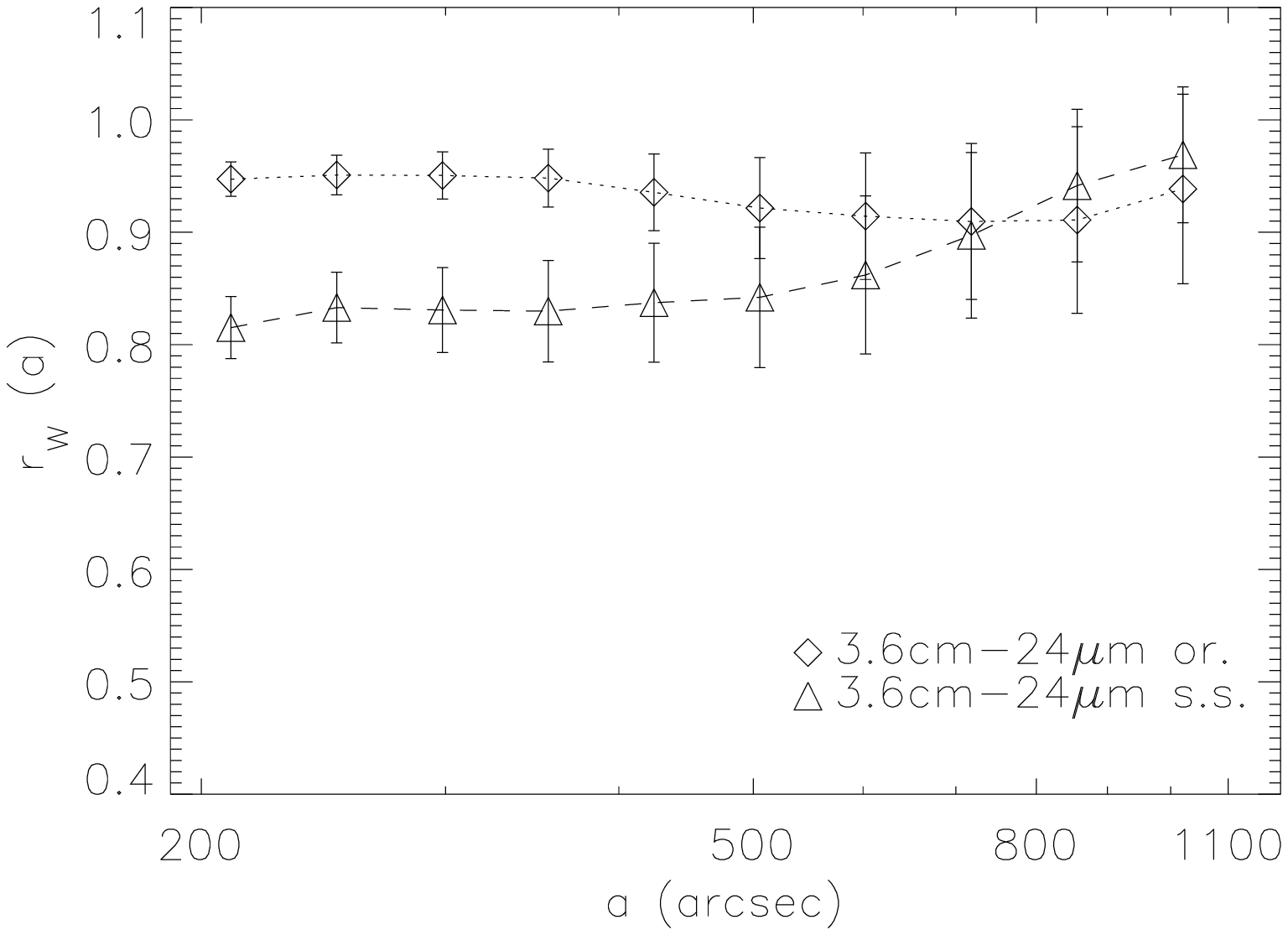}}
\resizebox{\hsize}{!}{\includegraphics*{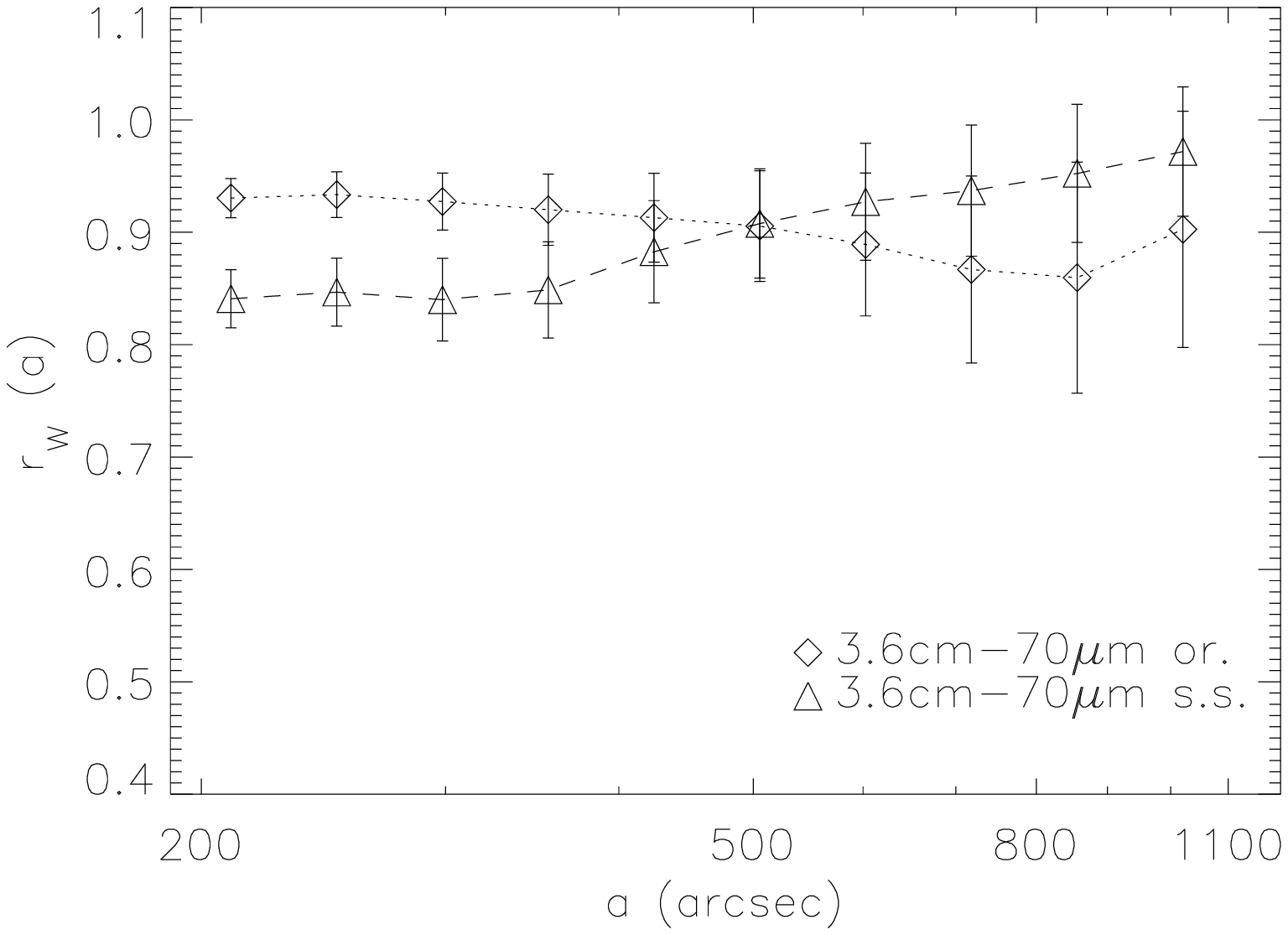}}
\resizebox{\hsize}{!}{\includegraphics*{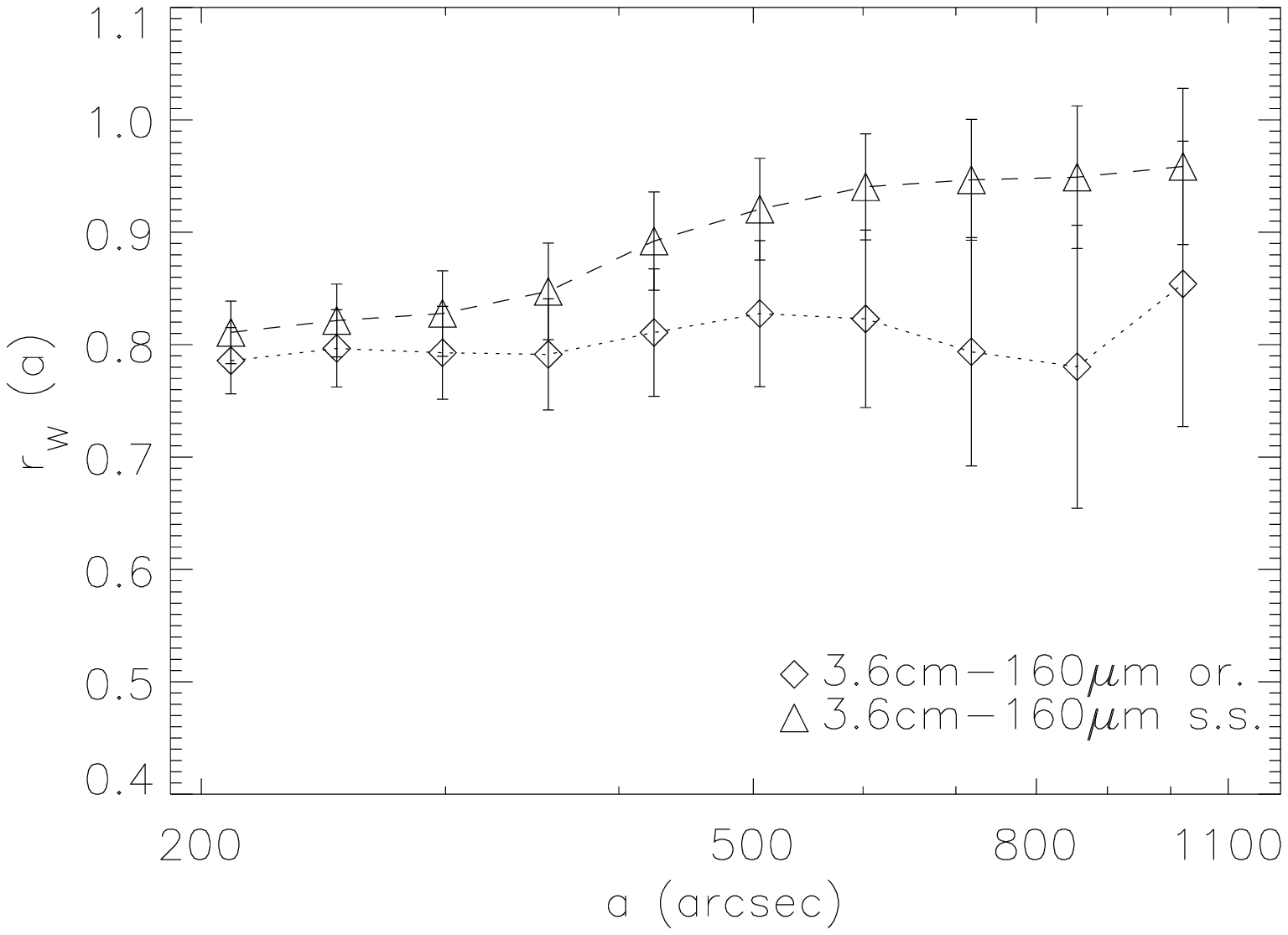}}
\caption[]{The cross--correlation between 3.6\,cm radio and IR before and after subtraction of the same sources at   $84\arcsec$ resolution.}
\end{figure}

\begin{figure}
\resizebox{\hsize}{!}{\includegraphics*{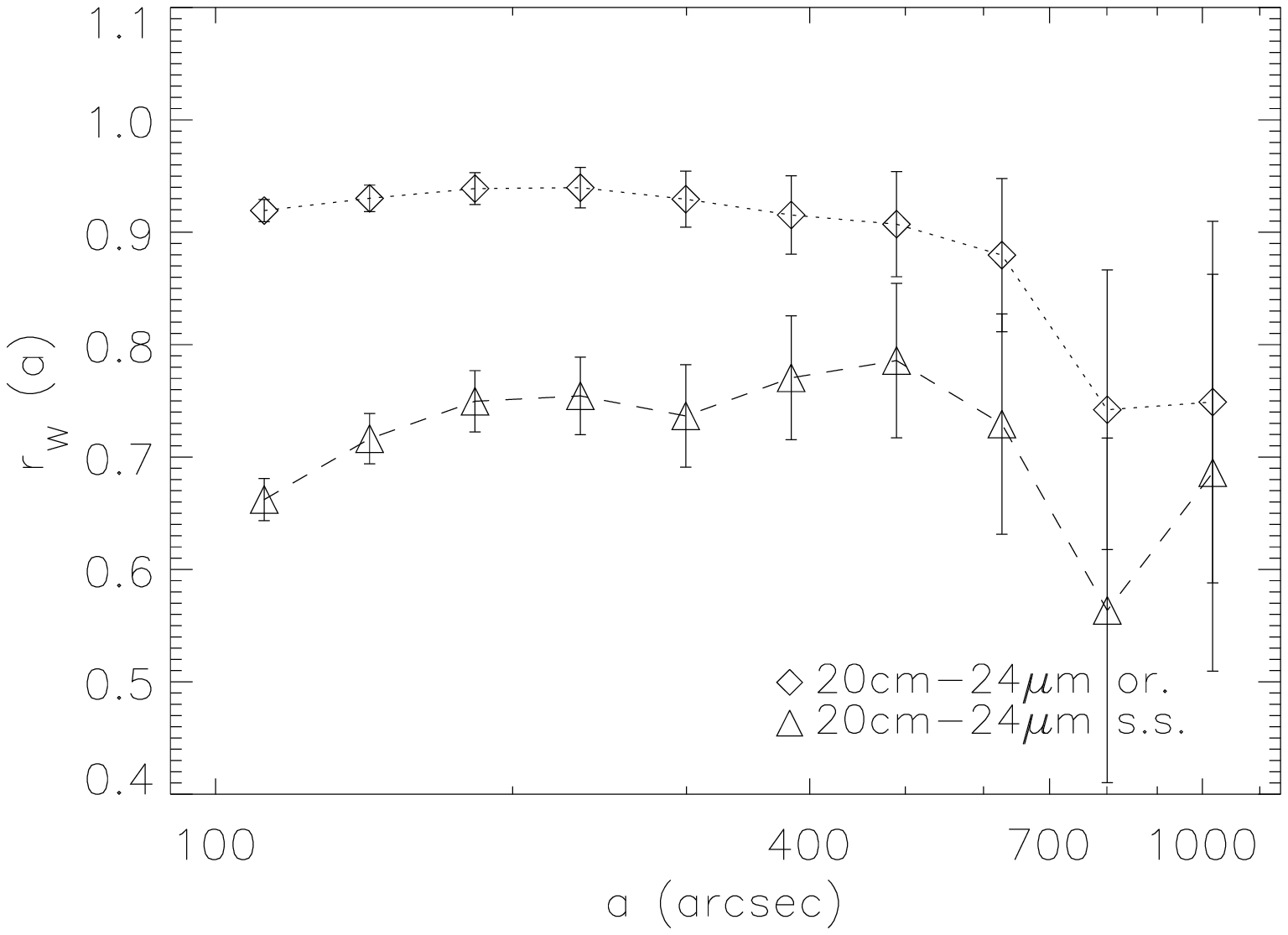}}
\resizebox{\hsize}{!}{\includegraphics*{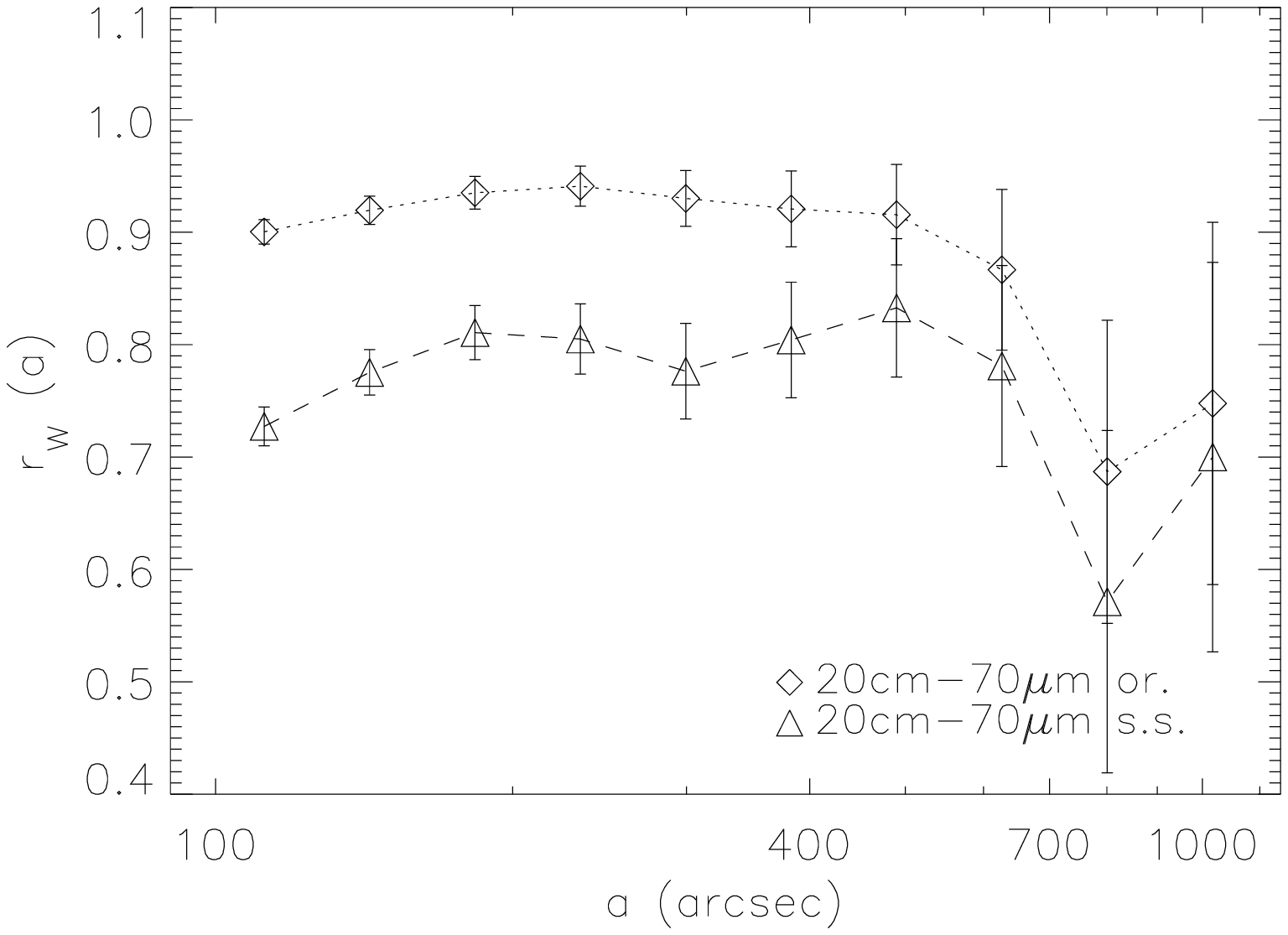}}
\resizebox{\hsize}{!}{\includegraphics*{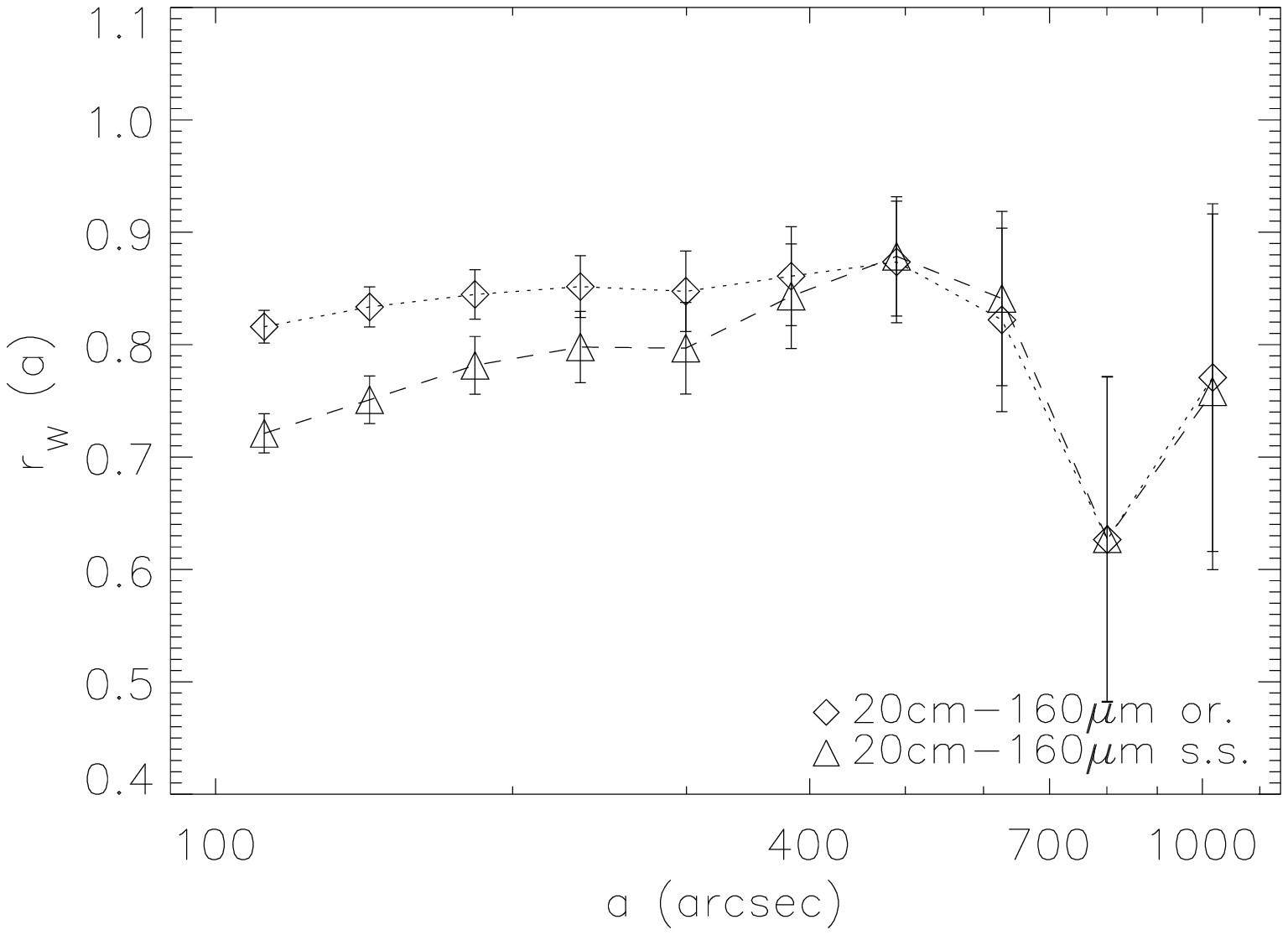}}
\caption[]{The correlation of the 20\,cm radio and IR before and after subtraction of the same sources at  $51\arcsec$ resolution (please note that this Fig. starts from 100$\arcsec$ and not 200$\arcsec$).}
\end{figure}

Second, we study the cross--correlations between the infrared and  the radio continuum emission at 3.6\,cm. 
As shown in Fig. 8, the correlations between 3.6\,cm radio and the three infrared emission are strong at all scales. At scales smaller than the width of the spiral arms (1.6\,kpc), the correlations are higher between 3.6\,cm and 24\,$\mu$m emission (and also 70\,$\mu$m) than between  3.6\,cm and 160\,$\mu$m emission. The source subtraction reduces the 3.6\,cm--24\,$\mu$m and 3.6\,cm--70\,$\mu$m correlations at scales smaller than the spiral arms. This indicates that young massive stars are the most common and important energy sources of the 3.6\,cm, 24 and 70\,$\mu$m emission at these scales.

Third, the cross--correlations with the radio continuum emission at 20\,cm are presented in Fig. 9. Before the source subtraction, the correlations between 20\,cm radio and the three IR bands are strong at scales smaller than the width of the spiral arms.  After the source subtraction, the coefficients of the IR--20\,cm correlations decrease dramatically so that they become less than 0.75 at small scales.
The reason is possibly that besides HII regions there are other unresolved sources of the 20\,cm radio emission like supernova remnants that are stronger than at 3.6\,cm but do not emit significantly in the infrared. Studies within our Galaxy have demonstrated that the ratio of far-infrared to radio continuum emission for supernova remnants is much smaller than that for HII regions~\citep{Fuerst}.

Comparing Fig. 8 with Fig. 9, a drop in correlation coefficient at the scale of $800\arcsec$ (3.2\,kpc) is more prominent in the MIPS correlations with 20\,cm than with 3.6\,cm. This is due to the strong minimum in the 20cm spectrum at this scale (Fig. 5) that persists even after the source subtraction (in contrast to the situation in the 3.6\,cm spectrum shown in Fig. 6).

\section{Comparison with H$\alpha$}

If we take the H$\alpha$ emission as a tracer of star forming regions that both heat the warm dust and ionize the gas giving the free--free emission, we expect a correlation between  H$\alpha$ emission, IR and thermal radio emission. The cross--correlations involving H$\alpha$ and IR images are shown in Fig. 10. The significant correlation of H$\alpha$ with the 24 and 70\,$\mu$m images at the smallest scale confirms  that most of the compact structures in the MIPS decomposed maps at 24 and 70\,$\mu$m (Fig. 3) correspond to the HII regions. The 160\,$\mu$m emission is also correlated with H$\alpha$ emission, although the correlations are weaker than the correlations at 24 and 70\,$\mu$m. It is deduced that the role of star forming regions in heating the dust, even the cold dust (emitting mainly at 160\,$\mu$m), is generally important at scales smaller than 4\,kpc.    

Between H$\alpha$ and 3.6\,cm there is a \emph{perfect correlation} for both small--scale and extended structures:  $r_{w}\geq$\,0.95 (bottom panel in Fig. 10).  
This is not surprising as the wavelet spectra of the  H$\alpha$ and 3.6\,cm maps are very similar (see Fig. 6). 
Thus, not only are the star forming regions mostly responsible for the 3.6\,cm emission, but also the radio continuum emission is dominated by the thermal free--free emission at 3.6\,cm for $a < 4$\,kpc.

After smoothing the 20\,cm map to the spatial resolution of the 3.6\,cm map, we obtain the H$\alpha$--20\,cm correlation (Fig. 10). At scales smaller than $600\arcsec$ or 2.4\,kpc, the coefficients are the same as for the H$\alpha$--3.6\,cm correlation. This means that the bright HII regions  are also important sources of the 20\,cm radio emission. 
It seems that the correlation decreases at larger scales, but it is not certain because of the large errors.

\begin{figure}
\resizebox{\hsize}{!}{\includegraphics*{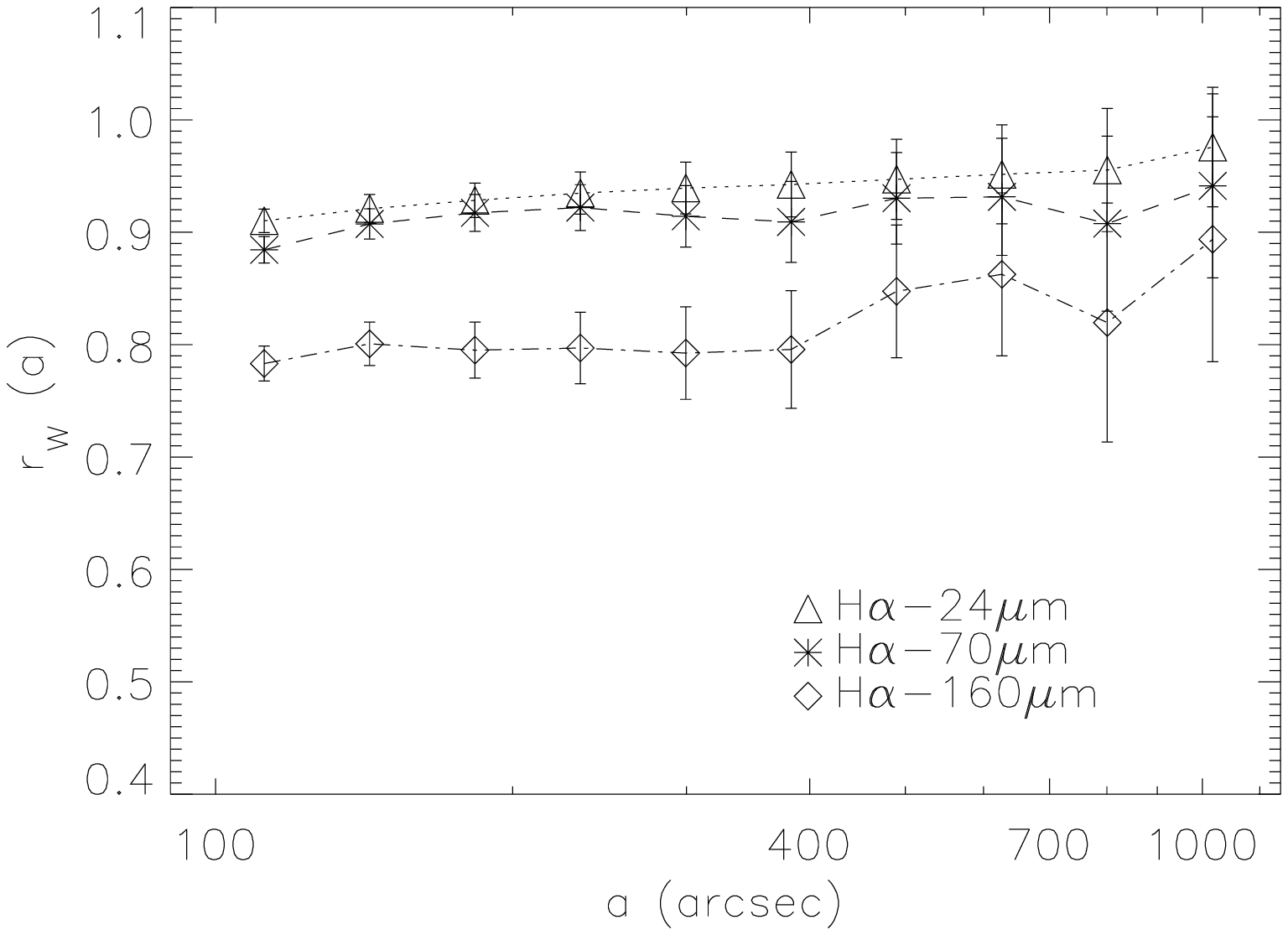}}
\resizebox{\hsize}{!}{\includegraphics*{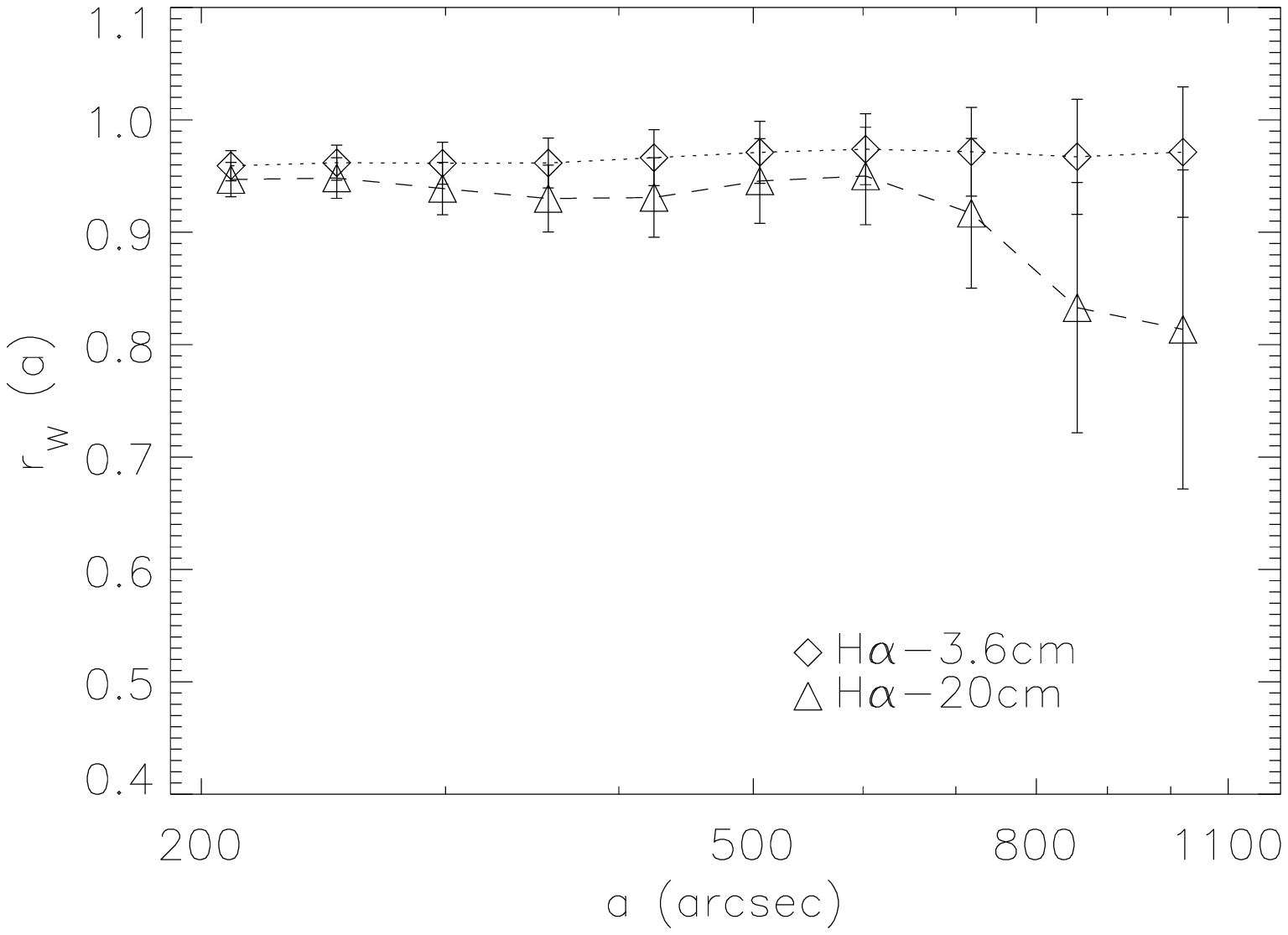}}
\caption[]{The correlation of H$\alpha$ with IR (top) at $51\arcsec$ and with 3.6 and 20\,cm radio (bottom) at $84\arcsec$ resolution. The maps include all HII regions.}
\end{figure}

\section{Discussion}

We studied the spectral characteristics of the MIPS mid-- and far--infrared, radio and H$\alpha$ maps and  obtained cross--correlation coefficients between pairs of maps. Here, we discuss the implications of our results for the understanding of the energy sources powering the IR and radio emission from M33. Further, we compare the results of our analysis on M33 with that of \cite{Frick_etal_01} on NGC\,6946. 

\subsection{IR emission}

From the wavelet analysis of the MIPS IR maps the following results are found.

\begin{itemize}
\item[$\bullet$] The 24 and 70\,$\mu$m emission emerge mostly from the bright sources corresponding to star forming and HII regions. The influence of HII regions on the IR emission is highest at 24\,$\mu$m (see Table 3).  As expected, at small scales the 24\,$\mu$m map is better correlated with the 70\,$\mu$m than with the 160\,$\mu$m map.

\item[$\bullet$] The 160\,$\mu$m emission emerges from both compact and extended structures. The compact structures  correspond to star forming and HII regions. As shown in Table 3, the fraction of wavelet energy provided by the 11 brightest HII regions at 160\,$\mu$m varies between 1/2 and 1/3 of that at 24\,$\mu$m at different scales. The 160\,$\mu$m map does not show small--scale dominance in its spectrum. The 160\,$\mu$m spectrum shows an increase when it reaches the scale of complexes of dust and gas clouds of $\sim$\,1 kpc, and a second increase in transition to the large--scale structures or diffuse dust emission.

\end{itemize}
While it is clear that the ionizing stars heat the warm dust, the mechanism heating the cold dust is still in debate.
According to our wavelet analysis, the direct role of the young stellar population decreases with increasing IR wavelength. As shown in Table 3, at small scales ($<$\,$200\arcsec$ or 0.8\,kpc), the relative contribution of the 11 brightest HII regions to the total emission energy is about 50\,$\%$ at 160\,$\mu$m which means that the contribution of all HII regions of M33 is much larger than 50$\%$. Hence, it seems that at least up to scales of 0.8\,kpc the cold dust is effectively heated by UV photons from massive ionizing stars. On the other hand, the 160\,$\mu$m--H$\alpha$ correlation coefficients are smaller than those of the 24\,$\mu$m~(or 70\,$\mu$m)--H$\alpha$ correlation (Fig. 10). This means that the number of structures with a specific scale emitting strongly in both IR and H$\alpha$ becomes less at 160\,$\mu$m. This indicates that besides UV photons from massive ionizing stars, other heating sources contribute also to heating the cold dust. These could be non--ionizing UV photons either from intermediate--mass stars (5--20 $M_{\odot}$) \citep{Xu_90} or from HII regions \citep{Popescu_etal_00,Popescu_02,Misiriotis}, or optical photons from solar mass stars \citep{Xu_96,Bianchi_00}; they should contribute to an average radiation field as the 160\,$\mu$m  wavelet spectrum is smooth.

Our findings confirm the arguments of e. g. \cite{Hinz}, \cite{Devereux_etal_94b,Devereux_etal_96,Devereux_etal_97} and \cite{Jones_etal_02} that the IR is powered predominantly by young O/B stars, specifically at 24 and 70\,$\mu$m. This is not necessarily at variance or in accordance with the argument of \cite{Xu_96} that only 27$\%$ of the diffuse dust emission from M31 can be attributed to the heating by UV photons, because of two reasons. First, our wavelet study does not include diffuse structures with scales larger than 4\,kpc. Second, the late--type galaxy M33 is more densely populated with young stars than the earlier--type galaxy M31. This may mean that the relative role of young stars in heating the dust increases with later galaxy type \citep{Sauvage_etal_92}.  For a more precise comparison, it is necessary to accomplish wavelet analysis for other nearby galaxies. 

\cite{Hoernes_etal_98} showed that the cold dust emission in M31 has a weaker correlation with the thermal radio than with the nonthermal radio emission. Since the contribution of the thermal emission is higher at 3.6\,cm than at 20\,cm, one may expect a weaker correlation between the 160\,$\mu$m and 3.6\,cm emission. In M33, we did not find significant differences between the 160\,$\mu$m--3.6\,cm and  160\,$\mu$m--20\,cm correlations (Figs. 8 and 9) at scales of $200\arcsec$\,(0.8\,kpc)\,$<$\,$a$\,$<$\,$700\arcsec$\,(2.8\,kpc), even after subtracting strong thermal sources (bright HII regions). This could be due to the larger role of UV photons from O/B stars in heating the cold dust in M33 than in M31, as discussed in the previous paragraph. However, a better 160\,$\mu$m correlation  with the nonthermal than with the thermal radio is probable at the scale of the whole galaxy. For instance, \cite{Hinz} showed that the morphology of the smoothed 160\,$\mu$m emission from M33 is most similar to that of the radio emission at 17.4\,cm.

\subsection{Radio emission and radio--IR correlation}

From the wavelet analysis of the radio maps we obtain the following results for the 3.6 and 20\,cm emission from M33.  \begin{itemize}
\item[$\bullet$] The shape of the 3.6\,cm wavelet spectrum is very similar to that of the H$\alpha$ spectrum (Fig. 6) and, hence,  there is a perfect correlation  between 3.6\,cm and H$\alpha$ at all scales (Fig. 10). This indicates that not only the star forming regions are mostly responsible for the radio emission at 3.6\,cm, but also the thermal (free--free) emission is the dominant component at this wavelength. But it could also be that the nonthermal (synchrotron) emission correlates well with the thermal emission. Our recent estimation of the thermal fraction of the 3.6\,cm radio continuum emission gives $f_{th}\sim 50\%$ for the whole galaxy including the extended disk emission (Tabatabaei et al., in prep.).  As a higher percentage of the thermal fraction is expected at scales $\leq$\,4\,kpc, this result is consistent with our wavelet study. 
The nearly parallel 3.6\,cm and H$\alpha$ wavelet spectra indicates that extinction is scale--independent and has no significant effect on H$\alpha$ emitting structures.

\item[$\bullet$] The combined 20\,cm (VLA+Effelsberg) wavelet spectrum represents both compact and diffuse structures of thermal and nonthermal emission. The bright HII regions (or complexes) give strong contribution to the 20\,cm emission. The strong H$\alpha$--20\,cm correlation at scales smaller than the width of the central extended region ($\sim$\,2.5\,kpc) also indicates that the nonthermal emission correlates perfectly with the thermal emission at these scales.

\item[$\bullet$] Generally, there are strong 3.6\,cm--IR correlations at all scales in the range of 0.8--4\,kpc within M33 (Fig. 8). The 3.6\,cm correlations with 24 and 70\,$\mu$m  are better than that with 160\,$\mu$m before subtracting bright HII regions. However, these correlations become approximately the same after subtracting HII regions. That is because the 24 and 70\,$\mu$m emission are more influenced by the HII regions than the 160\,$\mu$m emission.

\item[$\bullet$] The strong 20\,cm--IR correlations are mainly due to the bright HII regions, as the correlations fall  after source subtraction (Fig. 9). Here, the weakest correlation is between the 20\,cm and the 24\,$\mu$m maps. This indicates that the nonthermal sources which are associated with star forming regions are not as effective as HII regions in heating the dust. 
\end{itemize}
In their multi--resolution analysis of the radio--IR correlation in the LMC, \cite{Hughes_etal_06} found  strong  correlations in the regions where the thermal fraction of the radio emission is high. This is in agreement with our results  that the  24 and 70\,$\mu$m--radio correlations are  higher before subtracting HII regions. Even, the  160\,$\mu$m--20\,cm correlation follows this pattern before subtracting the thermal radio regions. For NGC6946, \cite{Frick_etal_01} also found a better correlation between the thermal 3.6\,cm radio and IR emission up to scales of the width of spiral arms, although the nonthermal emission dominates in this galaxy \citep{Ehle}

\subsection{Comparison  with NGC\,6946}

It is instructive to compare the results of the wavelet analysis of M33 with those obtained by \cite{Frick_etal_01} for NGC\,6946, which is also a late--type Scd galaxy. 

NGC\,6946 was studied at a linear resolution of 0.4\,kpc corresponding to our analysis at $84\arcsec$ HPBW. Comparing Fig. 6 (in this paper) to Fig. 7 in the \cite{Frick_etal_01} paper at scales between 0.8 and 4\,kpc, the dominant scale of the H$\alpha$ wavelet spectrum is at the scale of the width of the spiral arms (1.5\,kpc) in NGC\,6946, while in M33 it is at the smallest scale ($\sim$\,$200\arcsec$ or 0.8\,kpc in Fig. 6). Even after subtracting the 11 brightest HII regions  the scales corresponding to the spiral arms ($\sim$\,$400\arcsec$) are not dominant in M33.  This is because the spiral arms in M33 are not as pronounced as those in NGC\,6946. Instead, the  H$\alpha$ emission in M33 has a rather clumpy distribution.

The spectrum of dust emission from NGC\,6946 at 12--18\,$\mu$m shown in Fig. 7 in the \cite{Frick_etal_01} paper is more similar to the spectrum of dust emission from M33 at 160\,$\mu$m than to that at 24\,$\mu$m. So, the excess of hot dust emission near the massive stars found in M33 is not visible in NGC\,6946.

\cite{Frick_etal_01} found signatures of three--dimensional Kolmogorov--type  turbulence~\citep{Kolmogorov} for the ionized gas (H$\alpha$ and thermal radio continuum emission at 3.6\,cm) at scales less than 0.6\,kpc in NGC\,6946. An increasing part of the spectrum (logI--log$a$) with a slope of 5/3 is typical for this type of turbulence \citep{Spangler}.  They obtained this slope considering the first point at the scale of 0.3\,kpc (the resolution of their maps was 0.4\,kpc), making the slope estimate of 5/3 uncertain.  At the same resolution we cannot investigate this turbulence for M33, as the smallest scale in Fig. 6 is larger than 0.6\,kpc.
However, at higher spatial resolutions the spectra of the radio, IR, and H$\alpha$  emissions from M33 do not show such an increase at scales less than 0.6\,kpc (Figs. 4 and 5). The steepest increasing slope is about 1.1 at scales between 140 and 320\,pc (in the 24\,$\mu$m spectrum) (Fig. 4). The slopes become less at $51\arcsec$  resolution (Fig. 5). They vary between 0 and 0.6 at increasing parts of the spectra and between 0 and -0.9 at decreasing parts of the spectra. After subtracting HII regions, the slopes become even flatter.

Comparing Fig. 12 in the \cite{Frick_etal_01} paper to Fig. 10 (in this paper), the 3.6\,cm--H$\alpha$ cross--correlation coefficients are less at the small scales in NGC\,6946 than in M33. As the extinction is proportional to the dust density and because higher dust densities can be found at smaller scales, one expects more extinction of H$\alpha$  at smaller scales. This leads to a weaker correlation with the radio free--free emission at smaller scales. Hence, it seems that absorption by dust at scales $<$\,1.2\,kpc  \citep[Fig. 12, ][]{Frick_etal_01} is less significant in M33 than in NGC\,6946. The existence of considerable extinction in NGC\,6946 caused by a patchy dust distribution has been shown by \cite{Trewhella}.

\section{Summary}

Highly resolved and sensitive Spitzer MIPS images of M33 at 24, 70, and
160\,$\mu$m enabled us to compare the morphology of different dust components of this galaxy.
A 2D-wavelet transformation was used to find dominant scales of emitting structures and correlations of the MIPS images with new radio (3.6 and 20\,cm) maps as well as with an H$\alpha$ image of M33.
We compared the results of wavelet analysis with and without bright HII regions at different resolutions (18$\arcsec$, 51$\arcsec$, 84$\arcsec$).  We found that at a characteristic scale of $210\arcsec$ or 0.8\,kpc, 80\% or more of the wavelet energy at 24 and 70\,$\mu$m and 3.6\,cm is from the 11 brightest HII regions. These sources cause better correlations between pairs of the 24 and 70\,$\mu$m and 3.6\,cm maps, while they cause a smaller 24--160\,$\mu$m correlation.
Bright HII regions improve the correlations between the 20\,cm radio map and each of the MIPS IR maps. The H$\alpha$ emission shows a better correlation with IR emission at 24 and 70\,$\mu$m than at 160\,$\mu$m. It is perfectly correlated with the 3.6\,cm radio emission out to scales of 4\,kpc and also with the 20\,cm radio emission out to scales of the width of the central extended region ($\sim$\,2.5\,kpc).  This is understandable because the radio continuum emission at 3.6\,cm is dominated by thermal emission and that at 20\,cm by nonthermal emission. The most important conclusions of this study are:
\begin{itemize}
\item[$\bullet$] The longer the IR wavelength, the more extended is the distribution of dust grains emitting at that wavelength  (Sects. 3 and 4).

\item[$\bullet$] HII regions influence the IR emission with a strength inversely depending on wavelength: more influence at 24\,$\mu$m and less influence at 160\,$\mu$m (Sect. 4). 

\item[$\bullet$]  The nonthermal radio emission correlates well with the thermal radio and H$\alpha$ emission out to the scale of the central extended region, $\sim$\,2.5\,kpc (Sect. 7).

\item[$\bullet$] The warm dust--thermal radio correlation is stronger than the cold dust--nonthermal radio correlation at scales smaller than 4\,kpc (about half the size of the galaxy) (Sects. 5 and 7).

\item[$\bullet$] The effect of extinction in H$\alpha$ is independent of the scale of structures in M33 and is smaller than in  NGC6946.
\end{itemize}

\begin{acknowledgements}
We are grateful to R. Walterbos for kindly providing us with the H$\alpha$ data. 
We acknowledge NASA/SSC, University of Minnesota, and University of Arizona for the Spitzer MIPS data. We thank the staff of 100--m Effelsberg telescope and VLA  for their assistance with radio observations. The wavelet coefficients were calculated with the program WAVE2D which was written by Igor Patrikeev for MPIfR in 2001-2003. F. Tabatabaei was supported for this research through a stipend from the International Max Planck Research school (IMPRS) for Radio and Infrared Astronomy at the Universities of Bonn and Cologne. RDG, CEW, EP, KDG, JLH, and GHR were supported by NASA/JPL contracts 1256406, 1215746, and 1255094 to the Universities of Minnesota and Arizona. We thank P. L. Biermann, A. Fletcher, S. Willner,  P. Frick, and the anonymous referee for their helpful comments.
\end{acknowledgements}

\bibliography{s.bib}        

\end{document}